%% file: dynamics_channels_arxiv_v4.tex
\documentclass[10pt,showpacs,aps,twocolumn,nofootinbib,usenames,dvipsnames,svgnames,superscriptaddress, longbibliography]{revtex4-2}
\usepackage{xcolor, soul, color}
\definecolor{quantumviolet}{RGB}{79, 4, 134}
\definecolor{quantumgray}{RGB}{134,79,4}
\usepackage{enumitem}

\usepackage[linktocpage=true, colorlinks=true, linkcolor=blue, urlcolor=blue, citecolor=blue]{hyperref}

\input{preamble.tex}

\newcommand{\blk}{\color{black}}

\begin{document}
\title{
Revisiting dynamics of quantum causal structures -- when \emph{can} causal order 
evolve?}

\author{John H. Selby}
\email{john.h.selby@gmail.com}
\affiliation{International Centre for Theory of Quantum Technologies, University of Gda\'nsk, Jana Ba\.zy\'nskiego 1a, 80-309 Gda\'nsk, Poland}
\author{Ana Bel\'en Sainz}
\affiliation{International Centre for Theory of Quantum Technologies, University of Gda\'nsk, Jana Ba\.zy\'nskiego 1a, 80-309 Gda\'nsk, Poland}
\author{Pawe{\l} Horodecki}
\affiliation{International Centre for Theory of Quantum Technologies, University of Gda\'nsk, Jana Ba\.zy\'nskiego 1a, 80-309 Gda\'nsk, Poland}

\begin{abstract}
Recently, there has been substantial interest in studying the dynamics of quantum theory  beyond that of states, in particular, the dynamics of channels, measurements, and higher-order transformations.
{Ref.~[Phys.~Rev.~X 8(1), 011047 (2018)]} 
 pursues this using the process-matrix formalism, together with a definition of the possible dynamics of such process matrices, and focusing especially on the question of evolution of causal structures. One of its major conclusions is a strong theorem saying that within the formalism, under continuous and reversible transformations, the causal order between operations must be preserved. Our result here challenges that of {Ref.~[Phys.~Rev.~X 8(1), 011047 (2018)]}: if one is to take into account a full picture of the physical evolution of operations within the standard quantum--mechanical formalism, then the conclusion of {Ref.~[Phys.~Rev.~X 8(1), 011047 (2018)]} does not hold. That is, we show that under certain continuous and reversible dynamics, the causal order between operations is not necessarily preserved. We moreover identify and analyse the root of this apparent contradiction, specifically, that the commonly accepted and widely applied framework of higher-order processes, whilst mathematically sound, is not always appropriate for drawing conclusions on physical dynamics. Finally, we show how to reconcile the elements of the whole picture  following the intuition based on entanglement processing by local operations and classical communication.
\end{abstract}
\maketitle

\section{Introduction}

Quantum theory has, for over a century, challenged our \textit{common sense} in many scenarios---states of superpositions, uncertainty relations, and nonlocal phenomena, to name a few. One of its most recent impacts has been in the realm of causality: nowadays, one may theoretically construct scenarios where the cause--effect relations between objects is fundamentally undetermined as if in a quantum superposition. This concept was first introduced by Lucien Hardy in Ref.~\cite{hardy2005probability}, and further formalised in Refs.~\cite{hardy2007towards,markes2011entropy}, based on the idea that the interplay between quantum theory and general relativity seems to suggest that a unifying physical theory would feature a fundamentally indefinite structure of causal~\mbox{relations}.

It was first conjectured in Ref.~\cite{hardy2009quantum} that such indefiniteness should be a resource for information processing. Indeed, the seminal example of the so-called \textit{{quantum switch}},
 introduced in Ref.~\cite{chiribella2009beyond} and further developed in Ref.~\cite{qswitch}, shows that this is indeed the case. {There, an indefinite causal structure is simulated by using  quantum control of a causal structure,} that is, an auxiliary quantum system is used to determine whether a quantum process {$\mathcal{E}$} happens before or after another process {$\mathcal{F}$}---if the auxiliary system is initialised in a superposition state, then the causal structure between {$\mathcal{E}$} and {$\mathcal{F}$} is indefinite, which is also known as~causally non-separable \cite{branciard2016witnesses}. 
The quantum-switch experiment not only highlights new conceptual challenges for the foundations of physics but also presents an opportunity for quantum information theory, since it can be used to implement certain information-theoretic tasks that quantum circuits with definite causal structure cannot realise \cite{qswitch,araujo2014computational,feix2015quantum,guerin2016exponential,kristjansson2020resource,zhao2020quantum}. 

The question of indefinite causal structures is commonly tackled via the formalism of process matrices \cite{PM} ({it is worth} noting that there are alternative approaches to the study of indefinite causal structures, such as the aforementioned causaloid formaism of Hardy \cite{hardy2007towards}, the two-time formalism of Aharanov et~al.~\cite{aharonov1964time}, which was shown in Ref.~\cite{silva2017connecting} to subsume the process-matrix formalism, and the causally neutral formulation of quantum theory of Oreshkov and Cerf~\cite{oreshkov2016operational}.). In brief, a process matrix represents the most general way in which two local quantum operations occurring in two separate laboratories corresponding to two parties (Alice and Bob)
can be ``connected together''. There is, however, no predetermined causal order for the two quantum operations--- for example, we could have Alice's operation occurring prior to, after, or simultaneously to Bob's. Process matrices allow one to describe all of these situations but, moreover, also situations in which this causal order is (either classically or quantumly) indeterminate. That is, the formalism also allows for Alice and Bob to be connected together via a probabilistic mixture or even a quantum superposition of these definite causal orders.

This approach has since been substantially generalised to develop 
{an elaborate}, recursively defined, hierarchy of higher-order processes \cite{perinotti2017causal,kissinger2017categorical,bisio2019theoretical} building on earlier work on quantum combs \cite{chiribella2008quantum,chiribella2008transforming}. At the base of the hierarchy we have quantum states, then quantum channels (transforming states into states), {then} things that transform channels into channels, {then} process matrices, {then} things that transform process matrices into process matrices, and so on. This higher-order processes framework is highly relevant in various areas of physics: on the one hand, it provides the building blocks in certain mathematical frameworks for physical theories, such as Operational Probabilistic {Theories}  \cite{chiribella2010probabilistic}, and it also enables one to study physical situations where (for either fundamental or practical reasons)  the Schr\"odinger equation description cannot be applied. \blk

Recently, new interest has been {developing in}
the study not of quantum channels and process matrices themselves but rather on how these may change in time---i.e., their dynamics---for instance, over the course of an experiment. A recent work \cite{castro2018dynamics} uses the {aforementioned} formalism of higher-order processes to study the dynamics of process matrices. It was argued that if the dynamics are continuous and reversible; then, the causal order of the operations that the process matrix
acts on cannot be changed. That is, in order to set up, e.g., an experiment that features an  indefinite causal order starting from a process matrix whose causal structure is not indefinite, then one needs to implement a dynamical process that is discontinuous or irreversible. At first glance, this result appears to serve as a substantial roadblock toward using indefinite causal order as a resource for quantum information processing---after all, the world around us typically has a definite causal structure, so in order to use the indefinite causal structure as a resource, we will first need to generate it in some way. 

In this manuscript, however,  we present a different paradigm that takes place in the physics of dynamics of causal structures and helps overcome this roadblock. We discuss how an in-principle fine-grained description of the physical implementation of these processes helps one understand in which situations continuous and reversible transformations can change the causal structure of the process. In addition, we present a middle-ground description of the problem which helps analyse the time evolution of processes without taking a full fine-grained approach nor a full black-box approach. Before presenting these results, we will first overview the standard scenario and known results therein.

\section{Overview}\label{sec:overview}

In this paper, we revisit the subject of dynamics of causal orders and flesh out the apparent source of this roadblock. Indeed, we show how, at least in certain physical situations, this stems from semantics rather than fundamental physics. This paves the way for generating an  indefinite causal order as a resource for quantum information processing.

We will consider various objects whose dynamics we want to study:
states, channels, or process matrices.
We will moreover consider maps that act on these objects---these maps will correspond to the higher-order processes that {we will introduce more formally shortly.}
In the case of states, these maps are quantum channels, and in this manuscript, we will refer to them as S-maps.  In the case of channels, these maps are the so-called `clamps', and in this manuscript, we will refer to them as C-maps. In  the case of process matrices, these maps are the so-called `process-matrix transformations', and in this manuscript, we will refer to them as PM-maps.

To begin, let us revisit what it means for one of these objects to dynamically evolve~\cite{castro2018dynamics}. {In order to gain an intuition for the dynamics that process matrices can undergo, Ref.~\cite{castro2018dynamics} {notes} that quantum channels are a special case} of process matrices. {More} formally, {there is a (nonunique)} embedding of the set of channels into the set of process matrices. The process matrices in this embedding, however, behave differently under composition than the original channels do---for example, they can no longer be composed via tensor products \cite{jia2018tensor,guerin2019composition}. Ref.~\cite{castro2018dynamics} leverages the connection between quantum channels and process matrices to draw intuition {regarding} what the dynamics of a process matrix could possible be:

\begin{quotation}
   \it
   \noindent ``[...] quantum channels can evolve in time when the physical constituents employed for their implementation change.

\noindent    For example, the situation of a quantum system traversing an optical fibre whose properties change in time can be modelled by a time-evolving channel. The type of transformations we consider here can be seen as a generalisation [...] to the situation where the causal structure is indefinite''.
\end{quotation}

{The analysis of Ref.~\cite{castro2018dynamics} follows the intuition borrowed from the  standard theory of processes matrices and models the above evolution in time as a higher-order process on the type of object being evolved---an S-map makes a state evolve, a C-map makes a channel evolve, and a PM-map makes a process matrix evolve. Using this perspective, a strong mathematical theorem is proven \cite{castro2018dynamics} showing that continuous and reversible PM-maps are of a particular form which cannot change the causal order.}

Here, we show, {however,} that
when this formalism of higher-order processes is used to model the dynamics of a process matrix or a quantum channel, one may no longer capture some fundamental aspects of the full physical picture portrayed by the above quotation (though higher-order processes may still capture fundamental aspects of other physical situations). Indeed, we present several examples of physically realisable continuous and reversible dynamical evolution, which through the lens of higher-order processes however mathematically correspond to discontinuous or irreversible processes. As we will demonstrate, this can be easily seen just at the level of the dynamics of channels. In this manuscript, we {therefore} begin by discussing the case of the dynamics of quantum channels, presenting the example of the broken or vibrating fibre.
{We then move on to showing that} such {continuous and reversible} dynamics for process matrices can indeed change the causal order of the local operations. {Indeed,} we provide an example of evolution between definite causal orders as well as evolution between definite and indefinite orders.

Faced with this seeming contradiction between our examples and the results of Ref.~\cite{castro2018dynamics}---which as we mentioned earlier are mathematically sound---we are forced to reexamine the basic framework of higher-order processes itself. 
In doing so, we find that this framework is too limited to accommodate in full generality all physical situations of interest in a way that fundamental claims on the laws of physics can be always made. In light of this, we argue that  any physical conclusions regarding dynamics that are drawn based on the framework of higher-order processes are of limited applicability.  ({We} note that, as mentioned in the introduction, there are situations in which the framework of higher-order processes is the only one available to reason about the physical situations, and hence it should not always be disregarded.)

Simply stated, our argument can be captured already at the level of state preparations in the following way. The situation of a quantum system being prepared by some apparatus whose properties change in time can be modelled by a time-evolving density matrix. Now, if we ask how the reversible or continuous evolution of the apparatus looks from the perspective of the time-evolution of the associated density matrix, then we will see that the reversibility and continuity of the physical degrees of freedom of the apparatus do not imply the reversibility and continuity of the evolution at the level of the density matrix. This highlights the fact that unitary evolution tells us how a given state should evolve once it has been prepared by some physical apparatus; it does not tell us how the state that is prepared by some physical apparatus should change under the physical evolution of the apparatus itself.

{In this work, we further analyse} the assumptions underpinning the formalism of higher-order processes {and show in} which circumstances it {is an appropriate} formalism to use. 
In cases in which it is not, such as in the examples we present, then there is a need for a more {adequate} formalism for describing \blk dynamics. 
Towards this goal, here, we introduce a formalism for describing dynamics of channels, which {features a notion of evolution tailored at unraveling the fundamental (in terms of a more primitive object) \blk nature of channel dynamics. We} leave as a challenge for future work the development of an equivalent formalism for process matrices.

{It is worth mentioning that, taking a resource theoretic perspective, one may reconcile the results of Ref.~\cite{castro2018dynamics} with the ones presented in this paper---the difference merely lies in which kind of restrictions one wishes to impose in the set of allowed operations. A loose analogy to this may be drawn from entanglement theory \cite{horodecki2009quantum}: on the one hand, one can consider LO operations (local operations without communication), and on the other, one can consider LOCC operations (local operations with classical communication). Entanglement theory under LO operations is not a trivial one, since these operations allow for interesting manipulations of resources such as entanglement concentration within the pure states set. However, it is also known that LO is a much more restrictive set of operations than LOCC, and if one was to consider entanglement theory under LOCC operations, a broader scope of possibilities flourishes, such as the distillation of entanglement from mixed states. Coming back to the question of pertinence to this manuscript, we can view the results of Ref.~\cite{castro2018dynamics} as a non-trivial but restricted set of the types of transformations (and dynamics) that objects may undergo (the analogue of LO). Meanwhile, a more detailed control of the physics of the objects may lead to more general dynamics, as we present here (the analogue of LOCC), where more powerful tasks can be performed.}
\section{Channel Dynamics}

{Let us recall the basics of the mathematical formalism of higher-order processes that is popular in the field.
This formalism 
models} dynamics as being a particular class of linear maps from the set of `the objects being transformed' to itself. In the case where the objects being evolved are channels, the C-maps are represented by this particular `clamp'-shaped transformation $\mathcal{A}$:
\beq
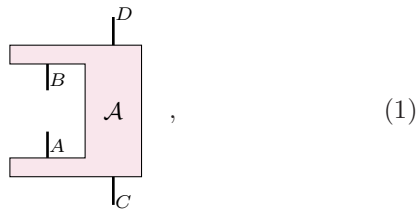
\begin{tikzpicture}
	\begin{pgfonlayer}{nodelayer}
		\node [style=none] (0) at (0, .5) {};
		\node [style=none] (1) at (0, 1.25) {};
		\node [style=none] (2) at (0, -1.25) {};
		\node [style={right label}] (3) at (0, -1) {$A$};
		\node [style={right label}] (4) at (0, 0.7500001) {$B$};
		\node [style=none] (5) at (-1, 1.25) {};
		\node [style=none] (6) at (-1, 1.75) {};
		\node [style=none] (7) at (1, 1.25) {};
		\node [style=none] (8) at (1, -1.25) {};
		\node [style=none] (9) at (-1, -1.25) {};
		\node [style=none] (10) at (-1, -1.75) {};
		\node [style=none] (11) at (2.5, -1.75) {};
		\node [style=none] (12) at (2.5, 1.75) {};
		\node [style=none] (13) at (1.75, -0) {\small $\mathcal{A}$};
		\node [style={right label}] (14) at (1.75, -2.5) {$C$};
		\node [style=none] (15) at (1.75, -1.75) {};
		\node [style=none] (16) at (1.75, -2.5) {};
		\node [style=none] (17) at (1.75, 2.5) {};
		\node [style={right label}] (18) at (1.75, 2.5) {$D$};
		\node [style=none] (19) at (1.75, 1.75) {};
		\node [style=none] (20) at (0, -.5) {};
	\end{pgfonlayer}
	\begin{pgfonlayer}{edgelayer}
		\draw [qWire] (2.center) to (20);
		\draw [qWire] (0) to (1.center);
		\filldraw[fill=purple!10,draw=black] (5.center) to (7.center) to (8.center) to (9.center)to (10.center) to (11.center) to (12.center) to (6.center) to cycle;
		\draw [qWire] (16.center) to (15.center);
		\draw [qWire] (19.center) to (17.center);
	\end{pgfonlayer}
\end{tikzpicture}\quad ,
\eeq
{where} 
 $A$, $B$, $C$, and $D$ are quantum systems. 

These C-maps can be formally defined as linear maps on the space of `channels from $A$ to $B$' to the space of `channels from $C$ to $D$', satisfying a generalised notion of complete positivity---when applied to one half of any bipartite channel, the result is another bipartite channel. That is, that for any bipartite channel $\mathcal{E}:X\otimes A \to Y\otimes B$, it must hold that the object defined {by}

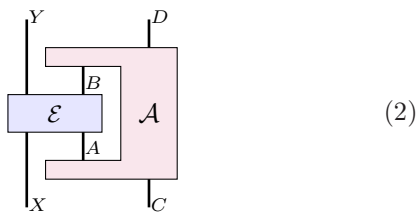
\beq
\begin{tikzpicture}
	\begin{pgfonlayer}{nodelayer}
		\node [style=none] (0) at (0, .45) {};
		\node [style=none] (1) at (0, 1.25) {};
		\node [style=none] (2) at (0, -1.25) {};
		\node [style={right label}] (3) at (0, -1) {$A$};
		\node [style={right label}] (4) at (0, 0.7500001) {$B$};
		\node [style=none] (5) at (-1, 1.25) {};
		\node [style=none] (6) at (-1, 1.75) {};
		\node [style=none] (7) at (1, 1.25) {};
		\node [style=none] (8) at (1, -1.25) {};
		\node [style=none] (9) at (-1, -1.25) {};
		\node [style=none] (10) at (-1, -1.75) {};
		\node [style=none] (11) at (2.5, -1.75) {};
		\node [style=none] (12) at (2.5, 1.75) {};
		\node [style=none] (13) at (1.75, -0) {$\mathcal{A}$};
		\node [style={right label}] (14) at (1.75, -2.5) {$C$};
		\node [style=none] (15) at (1.75, -1.75) {};
		\node [style=none] (16) at (1.75, -2.5) {};
		\node [style=none] (17) at (1.75, 2.5) {};
		\node [style={right label}] (18) at (1.75, 2.5) {$D$};
		\node [style=none] (19) at (1.75, 1.75) {};
		\node [style=none] (20) at (0, -.45) {};
		\node[style=none] (21) at (-2,-.5){};
		\node[style=none] (22) at (-2,.5){};
		\node[style=none] (23) at (.5,.5){};
		\node[style=none] (24) at (.5,-.5){};
		\node[style=none] (25) at (-.75,0){\small $\mathcal{E}$};
		\node[style=none](26) at (-1.5,-.45){};
		\node[style=none](27) at (-1.5,-2.5){};
		\node[style=none](28) at (-1.5,.45){};
		\node[style=none](29) at (-1.5,2.5){};
		\node [style={right label}] (30) at (-1.5, 2.5) {$Y$};
		\node [style={right label}] (31) at (-1.5, -2.5) {$X$};
	\end{pgfonlayer}
	\begin{pgfonlayer}{edgelayer}
			\filldraw[fill=purple!10,draw=black] (5.center) to (7.center) to (8.center) to (9.center)to (10.center) to (11.center) to (12.center) to (6.center) to cycle;
		\filldraw[fill=blue!10,draw=black] (21.center) to (22.center) to (23.center) to (24.center) to cycle;
		\draw [qWire] (29.center) to (28);
		\draw [qWire] (26) to (27.center);
		\draw [qWire] (2.center) to (20);
		\draw [qWire] (0) to (1.center);
		\draw [qWire] (16.center) to (15.center);
		\draw [qWire] (19.center) to (17.center);
	\end{pgfonlayer}
\end{tikzpicture}
\eeq
is a channel from $X\otimes C$ to $Y\otimes D$.

Now, the assumption made in {the structure theorem of} Ref.~\cite{castro2018dynamics} is that any dynamics of a channel can be expressed by a one-parameter family of such C-maps. That is, for any time-evolving channel $\mathcal{E}(t)$, we can always find C-maps $\mathcal{A}(t)$ such that the following equation is {satisfied}:
\beq
\begin{tikzpicture}\label{eq:belf}
	\begin{pgfonlayer}{nodelayer}
		\node [style={small box},fill=blue!10] (0) at (0, -0) {\small $\mathcal{E}(t)$};
		\node [style=none] (1) at (0, 1) {};
		\node [style=none] (2) at (0, -1) {};
		\node [style=right label] (3) at (0, -1) {$A$};
		\node [style=right label] (4) at (0, 1) {$B$};
	\end{pgfonlayer}
	\begin{pgfonlayer}{edgelayer}
		\draw [qWire] (2.center) to (0);
		\draw [qWire] (0) to (1.center);
	\end{pgfonlayer}
\end{tikzpicture}
\quad = \quad
\begin{tikzpicture}
	\begin{pgfonlayer}{nodelayer}
		\node [style={small box},fill=blue!10] (0) at (0, -0) {\small $\mathcal{E}(0)$};
		\node [style=none] (1) at (0, 1.25) {};
		\node [style=none] (2) at (0, -1.25) {};
		\node [style={right label}] (3) at (0, -1) {$A$};
		\node [style={right label}] (4) at (0, 0.7500001) {$B$};
		\node [style=none] (5) at (-1, 1.25) {};
		\node [style=none] (6) at (-1, 1.75) {};
		\node [style=none] (7) at (1, 1.25) {};
		\node [style=none] (8) at (1, -1.25) {};
		\node [style=none] (9) at (-1, -1.25) {};
		\node [style=none] (10) at (-1, -1.75) {};
		\node [style=none] (11) at (2.5, -1.75) {};
		\node [style=none] (12) at (2.5, 1.75) {};
		\node [style=none] (13) at (1.75, -0) {\small $\mathcal{A}(t)$};
		\node [style={right label}] (14) at (1.75, -2.5) {$A$};
		\node [style=none] (15) at (1.75, -1.75) {};
		\node [style=none] (16) at (1.75, -2.5) {};
		\node [style=none] (17) at (1.75, 2.5) {};
		\node [style={right label}] (18) at (1.75, 2.5) {$B$};
		\node [style=none] (19) at (1.75, 1.75) {};
	\end{pgfonlayer}
	\begin{pgfonlayer}{edgelayer}
			\filldraw[fill=purple!10,draw=black] (5.center) to (7.center) to (8.center) to (9.center)to (10.center) to (11.center) to (12.center) to (6.center) to cycle;
		\draw [qWire] (2.center) to (0);
		\draw [qWire] (0) to (1.center);
		\draw (5.center) to (7.center);
		\draw (7.center) to (8.center);
		\draw (8.center) to (9.center);
		\draw (9.center) to (10.center);
		\draw (11.center) to (10.center);
		\draw (11.center) to (12.center);
		\draw (12.center) to (6.center);
		\draw (6.center) to (5.center);
		\draw [qWire] (16.center) to (15.center);
		\draw [qWire] (19.center) to (17.center);
	\end{pgfonlayer}
\end{tikzpicture}\,.
\eeq

The types of dynamics captured by the mathematical expression of Equation~\eqref{eq:belf}, although applicable in certain physical situations, have however certain limitations, as we will show next. More precisely, we will show that the  sorts of dynamics discussed in the quote from Ref.~\cite{castro2018dynamics} mentioned above---a \emph{{``channel can evolve in time when the physical constituents employed for their implementation change.''}}---may not be fully covered by C-maps as per Equation~\eqref{eq:belf}. That is, when one `opens the box' and looks at the particular physical implementation of a channel, {one may appreciate fundamental fine-grained properties of nature that cannot be captured under} the above-mentioned assumption. 

First, we will argue that the natural quantum mechanical description of such evolution would not be written down as a higher-order process but instead as a description of how the physical constituents evolve according to the Schr\"odinger equation. Then, we will show that, typically, such evolution, {despite being fundamentally continuous and reversible,} cannot be understood as {driven by a continuous and reversible} higher-order process.

 {In later sections (Sections \ref{sec:Reexamined} and \ref{app:alternative}), we will moreover show that known attempts to accommodate these types of evolution within the framework of higher-order processes \cite{castro2018dynamics,castro2020comment}, even though successful, are not appropriate to always derive  fundamental claims---stemming from them---on the laws of physics.} 

\subsection{The broken fibre}

We prove our claim by presenting an extreme example based on the optical fibre discussed in the quote {from Ref.~\cite{castro2018dynamics}.} The particles which make up the physical constituents of the fibre will have a state described by some vector in an extremely large Hilbert space. (For the purpose of this paper, we only need the fact that in principle, one could write such a \textit{{microstate}}, 
 even though this is not achievable for any practical purposes. We do not need to consider any thermodynamical description of the fibre in terms of coarse-grained degrees of freedom. Therefore, we can leverage the reversibility of unitary quantum mechanics and do not encounter any irreversibilities arising from thermodynamic considerations.) To simplify this analysis, however, we will work within a 2D subspace spanned by two particular states of interest. These two states are `perfect', denoted by $\ket{\mathrm{p}}$, and `broken', denoted by $\ket{\mathrm{b}}$. The state $\ket{\mathrm{p}}$ describes the arrangement of the particles such that the fibre is working perfectly, that is, such that it implements the identity channel $\mathcal{I}$. In contrast, $\ket{\mathrm{b}}$ describes the arrangement of the particles where, for example, the fibre has broken in two, in which case the fibre would implement the totally depolarising channel $\mathcal{D}$. Now, there are clearly reversible dynamics which transform us continually and reversibly from one arrangement to the other.  ({{Notice} that here, we are considering a unitary evolution of a system which comprises all the constituents of the physical implementation of the fibre. The state of this system specifies all that is to know about its quantum mechanical description. We are never considering a thermodynamic description of the system with its corresponding macrostate.}) Consider the following Hamiltonian:
\beq
H = \ket{\mathrm{p}}\!\bra{\mathrm{b}} + \ket{\mathrm{b}}\!\bra{\mathrm{p}}\,,
\eeq
such that the time evolution of a state of the fibre is described by the unitary
\beq
U(t)=e^{itH}\,,
\eeq
which maps the state $\ket{\mathrm{p}}$ to $\ket{\mathrm{b}}$ in time $\pi/2$.

Let us now consider what this means for the description of the channel. At time $0$, we would describe the channel by the identity channel $\mathcal{I}$, whilst at time $t=\pi/2$, we would describe this as the totally depolarising channel $\mathcal{D}$. There is nothing particularly pathological about this example, save for how such a Hamiltonian describing the natural evolution of a system would be implemented.

The question then to ask is, can the continuous and reversible evolution driven by $U(t)$ be modelled as a continuous and reversible higher-order transformation on the space of channels?
The answer to this is no:  the identity channel is pure---Kraus rank $1$ and extremal in the convex set of channels---whilst the depolarising channel is mixed---maximal Kraus rank and  in the interior of the convex set of channels. Therefore, there cannot be a reversible C-map on the space of channels mapping one into the other.

\subsection{The vibrating fibre}

The clarity of the example of the broken fibre may be challenged by the mentioned implementation caveat. Here, we discuss the slightly more convoluted case of the vibrating fibre but for which there is a clear experimental realisation.

In order to demonstrate entanglement-enhanced classical communication, Ref.~\cite{Banaszek_et_al_2004} presents a practical demonstration of a depolarising channel with
finite time memory. Here, we will discuss how to adapt their setup such that the optical fibre can go from being an identity channel to being a depolarising channel, and back, through a physical setup that does not require the fibre to ``break into two''.

In Ref.~\cite{Banaszek_et_al_2004}, the essential idea is that the vibrations of the fibre change its birefringence randomly in a way that scrambles the polarisation degree of a single travelling photon. Therefore, in the scheme mentioned above, the optical fibre is connected to a sequence of coupled pendulums that stay excited during the whole experiment and {destabilise the fibre's polarisation activity in a continuous and steady way.} One can hence consider a new setup, where the pendulum systems become slowly excited and later on become de-excited due to dissipation continuously in time. This mechanism would affect differently in time the polarisation degrees of freedom of travelling photons, and the fibre will sometimes act as the identity channel and sometimes as a depolarising channel. In more abstract terms, the state of the environment causes the channel to be decohered and then to come back to its coherence state after some time.

This setup then implements the phenomenology that we are after: an optical fibre that can act as an identity or a depolarising channel and  can also be transformed from one state to the other via continuous and reversible transformations.

The only caveat that should be highlighted about this implementation is that, since our picture does not cover memory effects, subsequent uses of the channel should be timed so that the interval between the uses is long enough to go beyond the time-scale of correlations of the noise. In the particular experimental situation of Ref.~\cite{Banaszek_et_al_2004}, this interval would be certainly longer than 6 ns, since then the two consecutive photons experience the same perturbation, which is the essence of the improvement of classical communication.
We leave open for future research the questions of how superchannels with memory may be generally described and whether they preserve or not the causal structure of the channels they transform.

\section{Process matrix dynamics}\label{sec4}

The case where the objects being evolved are process matrices is a straightforward generalisation of the case of channel dynamics. {In Ref.~\citep{castro2018dynamics}, the assumption is made that} the evolution of process matrices should be described by higher-order processes on the space of process matrices, a PM-map, satisfying a generalised notion of complete positivity.

A process matrix is itself a higher-order process which we denote {as}
\beq
\begin{tikzpicture}\label{eq:thePM}
	\begin{pgfonlayer}{nodelayer}
		\node [style=none] (0) at (-1.75, 1.5) {};
		\node [style=none] (1) at (-1.75, 1) {};
		\node [style=none] (2) at (-0.75, 1) {};
		\node [style=none] (3) at (-0.75, -1) {};
		\node [style=none] (4) at (-1.75, -1) {};
		\node [style=none] (5) at (-1.75, -1.5) {};
		\node [style=none] (6) at (1.75, -1.5) {};
		\node [style=none] (7) at (1.75, -1) {};
		\node [style=none] (8) at (0.75, -1) {};
		\node [style=none] (9) at (0.75, 1) {};
		\node [style=none] (10) at (1.75, 1) {};
		\node [style=none] (11) at (1.75, 1.5) {};
		\node [style=none] (12) at (-1.5, 1) {};
		\node [style=none] (13) at (-1.5, 0.5) {};
		\node [style=none] (14) at (-1.5, -0.5) {};
		\node [style=none] (15) at (-1.5, -1) {};
		\node [style=none] (16) at (1.5, 1) {};
		\node [style=none] (17) at (1.5, 0.5) {};
		\node [style=none] (18) at (1.5, -0.5) {};
		\node [style=none] (19) at (1.5, -1) {};
		\node [style=none] (20) at (0, 0) {\small $W$};
		\node [style=none] (21) at (-1.5, -0.5) {};
		\node [style=none] (22) at (-1.5, -0.5) {};
		\node [style=right label] (23) at (-1.5, -0.675) {$A$};
		\node [style=right label] (24) at (-1.5, 0.5) {$B$};
		\node [style=right label] (25) at (.875, -0.675) {$C$};
		\node [style=right label] (26) at (.875, 0.5) {$D$};
	\end{pgfonlayer}
	\begin{pgfonlayer}{edgelayer}
		\filldraw[fill=blue!10,draw=black] (0.center) to (11.center) to (10.center) to (9.center) to (8.center) to (7.center) to (6.center) to (5.center) to (4.center) to (3.center) to (2.center) to (1.center)to cycle;
		\draw [style=qWire] (12.center) to (13.center);
		\draw [style=qWire] (14.center) to (15.center);
		\draw [style=qWire] (16.center) to (17.center);
		\draw [style=qWire] (18.center) to (19.center);
	\end{pgfonlayer}
\end{tikzpicture}
\eeq
which can be viewed as a map from a pair of channels (one `from $A$ to $B$', and one from `$C$ to $D$') to a probability that is subject to linearity and `complete positivity' conditions ({{the} precise nature of these conditions is not relevant here}). PM-maps are then defined as maps from process matrices to process matrices---a higher-order process $\mathcal{A}$ that we can depict {as}
\beq\label{eq:procMatrixTransf}
\begin{tikzpicture}
	\begin{pgfonlayer}{nodelayer}
		\node [style=none] (12) at (-1.5, 1) {};
		\node [style=none] (13) at (-1.5, 0.5) {};
		\node [style=none] (14) at (-1.5, -0.5) {};
		\node [style=none] (15) at (-1.5, -1) {};
		\node [style=none] (16) at (1.5, 1) {};
		\node [style=none] (17) at (1.5, 0.5) {};
		\node [style=none] (18) at (1.5, -0.5) {};
		\node [style=none] (19) at (1.5, -1) {};
		\node [style=none] (21) at (-1.5, -0.5) {};
		\node [style=none] (22) at (-1.5, -0.5) {};
		\node [style=right label] (23) at (-1.5, -1.125) {$A$};
		\node [style=right label] (24) at (-1.5, 1) {$B$};
		\node [style=right label] (25) at (.75, -1.125) {$C$};
		\node [style=right label] (26) at (.75, 1) {$D$};
		\node [style=none] (27) at (-1, -0.5) {};
		\node [style=none] (28) at (-1, 0.5) {};
		\node [style=none] (29) at (-2.25, 0.5) {};
		\node [style=none] (30) at (-2.25, -0.5) {};
		\node [style=none] (31) at (-2.25, -2) {};
		\node [style=none] (32) at (2.25, -2) {};
		\node [style=none] (33) at (2.25, -0.5) {};
		\node [style=none] (34) at (1, -0.5) {};
		\node [style=none] (35) at (1, 0.5) {};
		\node [style=none] (36) at (2.25, 0.5) {};
		\node [style=none] (37) at (2.25, 2) {};
		\node [style=none] (38) at (-2.25, 2) {};
		\node [style=none] (39) at (-4, 3) {};
		\node [style=none] (40) at (-4, 1) {};
		\node [style=none] (41) at (-3, 1) {};
		\node [style=none] (42) at (-3, -1) {};
		\node [style=none] (43) at (-4, -1) {};
		\node [style=none] (44) at (-4, -2.75) {};
		\node [style=none] (45) at (4, -2.75) {};
		\node [style=none] (46) at (4, -1) {};
		\node [style=none] (47) at (3, -1) {};
		\node [style=none] (48) at (3, 1) {};
		\node [style=none] (49) at (4, 1) {};
		\node [style=none] (50) at (4, 3) {};
		\node [style=none] (51) at (-3.75, 1) {};
		\node [style=none] (52) at (-3.75, 0.5) {};
		\node [style=none] (53) at (-3.75, -0.5) {};
		\node [style=none] (54) at (-3.75, -1) {};
		\node [style=none] (55) at (3.75, 1) {};
		\node [style=none] (56) at (3.75, 0.5) {};
		\node [style=none] (57) at (3.75, -0.5) {};
		\node [style=none] (58) at (3.75, -1) {};
		\node [style=none] (59) at (-3.75, -0.5) {};
		\node [style=none] (60) at (-3.75, -0.5) {};
		\node [style=right label] (61) at (-3.75, -0.675) {$A'$};
		\node [style=right label] (62) at (-3.75, 0.5) {$B'$};
		\node [style=right label] (63) at (3, -0.675) {$C'$};
		\node [style=right label] (64) at (3, 0.5) {$D'$};
		\node [style=none] (65) at (0, 2.5) {\small $\mathcal{A}$};
	\end{pgfonlayer}
	\begin{pgfonlayer}{edgelayer}
		\filldraw[fill=purple!10,draw=black] (39.center) to (50.center) to (49.center) to (48.center) to (47.center) to (46.center) to (45.center) to (44.center) to (43.center) to (42.center) to (41.center) to (40.center) to cycle;
		\filldraw[fill=white,draw=black] (38.center) to (37.center) to (36.center) to (35.center) to (34.center) to (33.center) to (32.center) to (31.center) to (30.center) to (27.center) to (28.center) to (29.center) to cycle;
		\draw [style=qWire] (51.center) to (52.center);
		\draw [style=qWire] (53.center) to (54.center);
		\draw [style=qWire] (55.center) to (56.center);
		\draw [style=qWire] (57.center) to (58.center);
		\draw [style=qWire] (12.center) to (13.center);
		\draw [style=qWire] (14.center) to (15.center);
		\draw [style=qWire] (16.center) to (17.center);
		\draw [style=qWire] (18.center) to (19.center);
	\end{pgfonlayer}
\end{tikzpicture}\,,
\eeq
where $A$, $B$, $C$, $D$, $A'$, $B'$, $C'$, and $D'$ are quantum systems.
The process $\mathcal{A}$ in Equation~\eqref{eq:procMatrixTransf} has a hole in the middle, into which we could plug in a process matrix on systems $A$, $B$, $C$ and $D$, which would give us a process matrix on systems $A'$, $B'$, $C'$ and $D'$. 

{The main result of Ref.~\cite{castro2018dynamics} focuses on the dynamics of process matrices that are represented as a one parameter family of such PM-maps, $\mathcal{A}(t)$, such {that}}
\beq\label{eq:theAoft}
\begin{tikzpicture}
	\begin{pgfonlayer}{nodelayer}
		\node [style=none] (0) at (-1.75, 1.5) {};
		\node [style=none] (1) at (-1.75, 1) {};
		\node [style=none] (2) at (-0.75, 1) {};
		\node [style=none] (3) at (-0.75, -1) {};
		\node [style=none] (4) at (-1.75, -1) {};
		\node [style=none] (5) at (-1.75, -1.5) {};
		\node [style=none] (6) at (1.75, -1.5) {};
		\node [style=none] (7) at (1.75, -1) {};
		\node [style=none] (8) at (0.75, -1) {};
		\node [style=none] (9) at (0.75, 1) {};
		\node [style=none] (10) at (1.75, 1) {};
		\node [style=none] (11) at (1.75, 1.5) {};
		\node [style=none] (12) at (-1.5, 1) {};
		\node [style=none] (13) at (-1.5, 0.5) {};
		\node [style=none] (14) at (-1.5, -0.5) {};
		\node [style=none] (15) at (-1.5, -1) {};
		\node [style=none] (16) at (1.5, 1) {};
		\node [style=none] (17) at (1.5, 0.5) {};
		\node [style=none] (18) at (1.5, -0.5) {};
		\node [style=none] (19) at (1.5, -1) {};
		\node [style=none] (20) at (0, 0) {\small $W(t)$};
		\node [style=none] (21) at (-1.5, -0.5) {};
		\node [style=none] (22) at (-1.5, -0.5) {};
		\node [style=right label] (23) at (-1.5, -0.675) {$A$};
		\node [style=right label] (24) at (-1.5, 0.5) {$B$};
		\node [style=right label] (25) at (.875, -0.675) {$C$};
		\node [style=right label] (26) at (.875, 0.5) {$D$};
	\end{pgfonlayer}
	\begin{pgfonlayer}{edgelayer}
			\filldraw[fill=blue!10,draw=black] (0.center) to (11.center) to (10.center) to (9.center) to (8.center) to (7.center) to (6.center) to (5.center) to (4.center) to (3.center) to (2.center) to (1.center)to cycle;
		\draw (0.center) to (11.center);
		\draw (11.center) to (10.center);
		\draw (10.center) to (9.center);
		\draw (9.center) to (8.center);
		\draw (8.center) to (7.center);
		\draw (7.center) to (6.center);
		\draw (6.center) to (5.center);
		\draw (5.center) to (4.center);
		\draw (4.center) to (3.center);
		\draw (3.center) to (2.center);
		\draw (2.center) to (1.center);
		\draw (1.center) to (0.center);
		\draw [style=qWire] (12.center) to (13.center);
		\draw [style=qWire] (14.center) to (15.center);
		\draw [style=qWire] (16.center) to (17.center);
		\draw [style=qWire] (18.center) to (19.center);
	\end{pgfonlayer}
\end{tikzpicture}
\quad = \quad
\begin{tikzpicture}
	\begin{pgfonlayer}{nodelayer}
		\node [style=none] (12) at (-1.5, 1) {};
		\node [style=none] (13) at (-1.5, 0.5) {};
		\node [style=none] (14) at (-1.5, -0.5) {};
		\node [style=none] (15) at (-1.5, -1) {};
		\node [style=none] (16) at (1.5, 1) {};
		\node [style=none] (17) at (1.5, 0.5) {};
		\node [style=none] (18) at (1.5, -0.5) {};
		\node [style=none] (19) at (1.5, -1) {};
		\node [style=none] (21) at (-1.5, -0.5) {};
		\node [style=none] (22) at (-1.5, -0.5) {};
		\node [style=right label] (23) at (-1.5, -0.875) {$A$};
		\node [style=right label] (24) at (-1.5, 0.675) {$B$};
		\node [style=right label] (25) at (0.75, -0.875) {$C$};
		\node [style=right label] (26) at (0.75, 0.675) {$D$};
		\node [style=none] (27) at (-1, -0.5) {};
		\node [style=none] (28) at (-1, 0.5) {};
		\node [style=none] (29) at (-2.25, 0.5) {};
		\node [style=none] (30) at (-2.25, -0.5) {};
		\node [style=none] (31) at (-2.25, -2) {};
		\node [style=none] (32) at (2.25, -2) {};
		\node [style=none] (33) at (2.25, -0.5) {};
		\node [style=none] (34) at (1, -0.5) {};
		\node [style=none] (35) at (1, 0.5) {};
		\node [style=none] (36) at (2.25, 0.5) {};
		\node [style=none] (37) at (2.25, 2) {};
		\node [style=none] (38) at (-2.25, 2) {};
		\node [style=none] (39) at (-4, 3) {};
		\node [style=none] (40) at (-4, 1) {};
		\node [style=none] (41) at (-3, 1) {};
		\node [style=none] (42) at (-3, -1) {};
		\node [style=none] (43) at (-4, -1) {};
		\node [style=none] (44) at (-4, -2.75) {};
		\node [style=none] (45) at (4, -2.75) {};
		\node [style=none] (46) at (4, -1) {};
		\node [style=none] (47) at (3, -1) {};
		\node [style=none] (48) at (3, 1) {};
		\node [style=none] (49) at (4, 1) {};
		\node [style=none] (50) at (4, 3) {};
		\node [style=none] (51) at (-3.75, 1) {};
		\node [style=none] (52) at (-3.75, 0.5) {};
		\node [style=none] (53) at (-3.75, -0.5) {};
		\node [style=none] (54) at (-3.75, -1) {};
		\node [style=none] (55) at (3.75, 1) {};
		\node [style=none] (56) at (3.75, 0.5) {};
		\node [style=none] (57) at (3.75, -0.5) {};
		\node [style=none] (58) at (3.75, -1) {};
		\node [style=none] (59) at (-3.75, -0.5) {};
		\node [style=none] (60) at (-3.75, -0.5) {};
		\node [style=right label] (61) at (-3.75, -0.675) {$A$};
		\node [style=right label] (62) at (-3.75, 0.5) {$B$};
		\node [style=right label] (63) at (3.125, -0.675) {$C$};
		\node [style=right label] (64) at (3.125, 0.5) {$D$};
		\node [style=none] (65) at (0, 2.5) {\small $\mathcal{A}(t)$};
		\node [style=none] (66) at (-1.75, 1) {};
		\node [style=none] (67) at (-1.75, 1.5) {};
		\node [style=none] (68) at (1.75, 1.5) {};
		\node [style=none] (69) at (1.75, 1) {};
		\node [style=none] (70) at (0.75, 1) {};
		\node [style=none] (71) at (-0.75, 1) {};
		\node [style=none] (72) at (-0.75, -1) {};
		\node [style=none] (73) at (-2, -1) {};
		\node [style=none] (74) at (-2, -1.5) {};
		\node [style=none] (75) at (2, -1.5) {};
		\node [style=none] (76) at (2, -1) {};
		\node [style=none] (77) at (0.75, -1) {};
		\node [style=none] (78) at (0, 0) {\small $W(0)$};
	\end{pgfonlayer}
	\begin{pgfonlayer}{edgelayer}
			\filldraw[fill=purple!10,draw=black] (39.center) to (50.center) to (49.center) to (48.center) to (47.center) to (46.center) to (45.center) to (44.center) to (43.center) to (42.center) to (41.center) to (40.center) to cycle;
		\filldraw[fill=white,draw=black] (38.center) to (37.center) to (36.center) to (35.center) to (34.center) to (33.center) to (32.center) to (31.center) to (30.center) to (27.center) to (28.center) to (29.center) to cycle;
		\draw [style=qWire] (12.center) to (13.center);
		\draw [style=qWire] (14.center) to (15.center);
		\draw [style=qWire] (16.center) to (17.center);
		\draw [style=qWire] (18.center) to (19.center);
		\draw [style=qWire] (51.center) to (52.center);
		\draw [style=qWire] (53.center) to (54.center);
		\draw [style=qWire] (55.center) to (56.center);
		\draw [style=qWire] (57.center) to (58.center);
		\filldraw[fill=blue!10,draw=black] (67.center) to (68.center) to (69.center) to (70.center) to (77.center) to (76.center) to (75.center) to (74.center) to (73.center) to (72.center) to (71.center) to (66.center) to cycle;
	\end{pgfonlayer}
\end{tikzpicture}\,.
\eeq
{In particular, Ref.~\cite{castro2018dynamics} proves a structure theorem for the transformations $\mathcal{A}(t)$:  those that are continuous and reversible, factor into four unitary transformations, one acting on each input and output of $W(0)$, {that is},}
\beq
\begin{tikzpicture}
	\begin{pgfonlayer}{nodelayer}
		\node [style=none] (12) at (-1.5, 1) {};
		\node [style=none] (13) at (-1.5, 0.5) {};
		\node [style=none] (14) at (-1.5, -0.5) {};
		\node [style=none] (15) at (-1.5, -1) {};
		\node [style=none] (16) at (1.5, 1) {};
		\node [style=none] (17) at (1.5, 0.5) {};
		\node [style=none] (18) at (1.5, -0.5) {};
		\node [style=none] (19) at (1.5, -1) {};
		\node [style=none] (21) at (-1.5, -0.5) {};
		\node [style=none] (22) at (-1.5, -0.5) {};
		\node [style=right label] (23) at (-1.5, -0.825) {$A$};
		\node [style=right label] (24) at (-1.5, 0.625) {$B$};
		\node [style=right label] (25) at (.875, -0.825) {$C$};
		\node [style=right label] (26) at (.875, 0.625) {$D$};
		\node [style=none] (27) at (-1, -0.5) {};
		\node [style=none] (28) at (-1, 0.5) {};
		\node [style=none] (29) at (-2.25, 0.5) {};
		\node [style=none] (30) at (-2.25, -0.5) {};
		\node [style=none] (31) at (-2.25, -2) {};
		\node [style=none] (32) at (2.25, -2) {};
		\node [style=none] (33) at (2.25, -0.5) {};
		\node [style=none] (34) at (1, -0.5) {};
		\node [style=none] (35) at (1, 0.5) {};
		\node [style=none] (36) at (2.25, 0.5) {};
		\node [style=none] (37) at (2.25, 2) {};
		\node [style=none] (38) at (-2.25, 2) {};
		\node [style=none] (39) at (-4, 3) {};
		\node [style=none] (40) at (-4, 1) {};
		\node [style=none] (41) at (-3, 1) {};
		\node [style=none] (42) at (-3, -1) {};
		\node [style=none] (43) at (-4, -1) {};
		\node [style=none] (44) at (-4, -2.75) {};
		\node [style=none] (45) at (4, -2.75) {};
		\node [style=none] (46) at (4, -1) {};
		\node [style=none] (47) at (3, -1) {};
		\node [style=none] (48) at (3, 1) {};
		\node [style=none] (49) at (4, 1) {};
		\node [style=none] (50) at (4, 3) {};
		\node [style=none] (51) at (-3.75, 1) {};
		\node [style=none] (52) at (-3.75, 0.5) {};
		\node [style=none] (53) at (-3.75, -0.5) {};
		\node [style=none] (54) at (-3.75, -1) {};
		\node [style=none] (55) at (3.75, 1) {};
		\node [style=none] (56) at (3.75, 0.5) {};
		\node [style=none] (57) at (3.75, -0.5) {};
		\node [style=none] (58) at (3.75, -1) {};
		\node [style=none] (59) at (-3.75, -0.5) {};
		\node [style=none] (60) at (-3.75, -0.5) {};
		\node [style=right label] (61) at (-3.75, -0.5) {$A$};
		\node [style=right label] (62) at (-3.75, 0.5) {$B$};
		\node [style=right label] (63) at (3.125, -0.5) {$C$};
		\node [style=right label] (64) at (3.125, 0.5) {$D$};
		\node [style=none] (65) at (0, 2.5) {$\mathcal{A}(t)$};
		\node [style=none] (66) at (-1.75, 1) {};
		\node [style=none] (67) at (-1.75, 1.5) {};
		\node [style=none] (68) at (1.75, 1.5) {};
		\node [style=none] (69) at (1.75, 1) {};
		\node [style=none] (70) at (0.75, 1) {};
		\node [style=none] (71) at (-0.75, 1) {};
		\node [style=none] (72) at (-0.75, -1) {};
		\node [style=none] (73) at (-2, -1) {};
		\node [style=none] (74) at (-2, -1.5) {};
		\node [style=none] (75) at (2, -1.5) {};
		\node [style=none] (76) at (2, -1) {};
		\node [style=none] (77) at (0.75, -1) {};
		\node [style=none] (78) at (0, 0) {\small $W(0)$};
	\end{pgfonlayer}
	\begin{pgfonlayer}{edgelayer}
			\filldraw[fill=purple!10,draw=black] (39.center) to (50.center) to (49.center) to (48.center) to (47.center) to (46.center) to (45.center) to (44.center) to (43.center) to (42.center) to (41.center) to (40.center) to cycle;
		\filldraw[fill=white,draw=black] (38.center) to (37.center) to (36.center) to (35.center) to (34.center) to (33.center) to (32.center) to (31.center) to (30.center) to (27.center) to (28.center) to (29.center) to cycle;
				\filldraw[fill=blue!10,draw=black] (67.center) to (68.center) to (69.center) to (70.center) to (77.center) to (76.center) to (75.center) to (74.center) to (73.center) to (72.center) to (71.center) to (66.center) to cycle;
		\draw [style=qWire] (12.center) to (13.center);
		\draw [style=qWire] (14.center) to (15.center);
		\draw [style=qWire] (16.center) to (17.center);
		\draw [style=qWire] (18.center) to (19.center);
		\draw [style=qWire] (51.center) to (52.center);
		\draw [style=qWire] (53.center) to (54.center);
		\draw [style=qWire] (55.center) to (56.center);
		\draw [style=qWire] (57.center) to (58.center);
	\end{pgfonlayer}
\end{tikzpicture}
\quad =\quad
\begin{tikzpicture}
	\begin{pgfonlayer}{nodelayer}
		\node [style=none] (12) at (-2, 2.75) {};
		\node [style=none] (13) at (-2, 2.25) {};
		\node [style=none] (14) at (-2, -2.25) {};
		\node [style=none] (15) at (-2, -2.75) {};
		\node [style=none] (16) at (2, 2.75) {};
		\node [style=none] (17) at (2, 2.25) {};
		\node [style=none] (18) at (2, -2.25) {};
		\node [style=none] (19) at (2, -2.75) {};
		\node [style=none] (21) at (-2, -2.25) {};
		\node [style=none] (22) at (-2, -2.25) {};
		\node [style=right label] (23) at (-2, -2.625) {$A$};
		\node [style=right label] (24) at (-2, 2.375) {$B$};
		\node [style=right label] (25) at (1.375, -2.625) {$C$};
		\node [style=right label] (26) at (1.375, 2.375) {$D$};
		\node [style=none] (66) at (-2.25, 2.75) {};
		\node [style=none] (67) at (-2.25, 3.25) {};
		\node [style=none] (68) at (2.25, 3.25) {};
		\node [style=none] (69) at (2.25, 2.75) {};
		\node [style=none] (70) at (0.75, 2.75) {};
		\node [style=none] (71) at (-0.75, 2.75) {};
		\node [style=none] (72) at (-0.75, -2.75) {};
		\node [style=none] (73) at (-2.25, -2.75) {};
		\node [style=none] (74) at (-2.25, -3.25) {};
		\node [style=none] (75) at (2.25, -3.25) {};
		\node [style=none] (76) at (2.25, -2.75) {};
		\node [style=none] (77) at (0.75, -2.75) {};
		\node [style=none] (78) at (0, 0) {\small $W(0)$};
		\node [style=small box, fill=purple!10] (79) at (-2, 1.75) {\small $U_B(t)$};
		\node [style=small box, fill=purple!10] (80) at (-2, -1.75) {\small $U_A(t)$};
		\node [style=small box, fill=purple!10] (81) at (2, 1.75) {\small $U_D(t)$};
		\node [style=small box, fill=purple!10] (82) at (2, -1.75) {\small $U_C(t)$};
		\node [style=none] (83) at (-2, 1.25) {};
		\node [style=none] (84) at (-2, 0.75) {};
		\node [style=right label] (85) at (-2, .75) {$B$};
		\node [style=none] (86) at (-2, -0.75) {};
		\node [style=none] (87) at (-2, -1.25) {};
		\node [style=none] (88) at (-2, -0.75) {};
		\node [style=none] (89) at (-2, -0.75) {};
		\node [style=right label] (90) at (-2, -1) {$A$};
		\node [style=none] (91) at (2, -0.75) {};
		\node [style=none] (92) at (2, -1.25) {};
		\node [style=right label] (93) at (1.375, -1) {$C$};
		\node [style=none] (94) at (0.75, -1.25) {};
		\node [style=none] (95) at (2, 1.25) {};
		\node [style=none] (96) at (2, 0.75) {};
		\node [style=right label] (97) at (1.375, .75) {$D$};
		\node [style=none] (98) at (0.75, 1.25) {};
	\end{pgfonlayer}
	\begin{pgfonlayer}{edgelayer}
		\draw [style=qWire] (12.center) to (13.center);
		\draw [style=qWire] (14.center) to (15.center);
		\draw [style=qWire] (16.center) to (17.center);
		\draw [style=qWire] (18.center) to (19.center);
		\filldraw[fill=blue!10,draw=black] (67.center) to (68.center) to (69.center) to (70.center) to (77.center) to (76.center) to (75.center) to (74.center) to (73.center) to (72.center) to (71.center) to (66.center) to cycle;
		\draw [style=qWire] (83.center) to (84.center);
		\draw [style=qWire] (86.center) to (87.center);
		\draw [style=qWire] (91.center) to (92.center);
		\draw [style=qWire] (95.center) to (96.center);
	\end{pgfonlayer}
\end{tikzpicture}\,.
\eeq
{From this result, a no-go theorem for the type of dynamics that can be driven by $\mathcal{A}(t)$ can be derived \cite{castro2018dynamics}:} {dynamics by a continuous and reversible $\mathcal{A}(t)$ cannot change the global causal ordering of $W(0)$,  as $\mathcal{A}(t)$ acts only locally on Alice and Bob's inputs and outputs. Finally, 
{the result is turned} into a no-go theorem for fundamental physical dynamics: the physical evolution that changes the global causal ordering of $W(0)$ must be fundamentally discontinuous or irreversible. Here, we argue that this physical no-go theorem does not really follow from the mathematical one (which we believe to be correct), and that there are instances where this physical no-go theorem indeed does not hold.}

We will now present two examples where continuous reversible dynamics of process matrices do indeed change the causal order. With this, we show that the fundamental properties of the dynamics of process matrices cannot be drawn exclusively from the features of such a family of higher-order transformation $\mathcal{A}(t)$.

\subsection{Rewiring: Dynamics between Definite Causal Orders}

In this first example, we will define two process matrices (which have different causal order) and show that there exists a continuous and reversible transformation between~them.

The first process matrix, following the notation of Equation~\eqref{eq:thePM}, consists of (i) the quantum system $A$ being prepared in a fixed state $\rho$, (ii) an identity channel $\mathcal{I}$ from the system $B$ to $C$, and (iii) the system $D$ being discarded ($\mathsf{tr}$). This process matrix has a causal influence from the left wing (hereon, Alice) to the right wing (hereon, Bob),  so we denote it as $W_{\mathbf{B}\to \mathbf{C}}$. Diagrammatically, this is depicted as follows:
\beq
\begin{tikzpicture}
	\begin{pgfonlayer}{nodelayer}
		\node [style=none] (0) at (-2.25, 1.5) {};
		\node [style=none] (1) at (-2.25, 1) {};
		\node [style=none] (2) at (-1.25, 1) {};
		\node [style=none] (3) at (-1.25, -1) {};
		\node [style=none] (4) at (-2.25, -1) {};
		\node [style=none] (5) at (-2.25, -1.5) {};
		\node [style=none] (6) at (2.25, -1.5) {};
		\node [style=none] (7) at (2.25, -1) {};
		\node [style=none] (8) at (1.25, -1) {};
		\node [style=none] (9) at (1.25, 1) {};
		\node [style=none] (10) at (2.25, 1) {};
		\node [style=none] (11) at (2.25, 1.5) {};
		\node [style=none] (12) at (-2, 1) {};
		\node [style=none] (13) at (-2, 0.5) {};
		\node [style=none] (14) at (-2, -0.5) {};
		\node [style=none] (15) at (-2, -1) {};
		\node [style=none] (16) at (2, 1) {};
		\node [style=none] (17) at (2, 0.5) {};
		\node [style=none] (18) at (2, -0.5) {};
		\node [style=none] (19) at (2, -1) {};
		\node [style=none] (20) at (0, 0) {\small $W_{\mathbf{B}\to \mathbf{C}}$};
		\node [style=none] (21) at (-2, -0.5) {};
		\node [style=none] (22) at (-2, -0.5) {};
		\node [style={right label}] (23) at (-2, -0.625) {$A$};
		\node [style={right label}] (24) at (-2, 0.5) {$B$};
		\node [style={right label}] (25) at (1.375, -0.625) {$C$};
		\node [style={right label}] (26) at (1.375, 0.5) {$D$};
	\end{pgfonlayer}
	\begin{pgfonlayer}{edgelayer}
		\filldraw[fill=blue!10,draw=black] (0.center) to (11.center) to (10.center) to (9.center) to (8.center) to (7.center) to (6.center) to (5.center) to (4.center) to (3.center) to (2.center) to (1.center)to cycle;
		\draw [style=qWire] (12.center) to (13.center);
		\draw [style=qWire] (14.center) to (15.center);
		\draw [style=qWire] (16.center) to (17.center);
		\draw [style=qWire] (18.center) to (19.center);
	\end{pgfonlayer}
\end{tikzpicture}
\quad =\quad
\begin{tikzpicture}
	\begin{pgfonlayer}{nodelayer}
		\node [style=none] (0) at (-2.25, 2) {};
		\node [style=none] (1) at (-2.25, .75) {};
		\node [style=none] (2) at (-0.75, .75) {};
		\node [style=none] (3) at (-0.75, -.75) {};
		\node [style=none] (4) at (-2.25, -.75) {};
		\node [style=none] (5) at (-2.25, -2) {};
		\node [style=none] (6) at (2.25, -2) {};
		\node [style=none] (7) at (2.25, -.75) {};
		\node [style=none] (8) at (0.75, -.75) {};
		\node [style=none] (9) at (0.75, .75) {};
		\node [style=none] (10) at (2.25, .75) {};
		\node [style=none] (11) at (2.25, 2) {};
		\node [style=none] (12) at (-1.5, .75) {};
		\node [style=none] (13) at (-1.5, 0.5) {};
		\node [style=none] (14) at (-1.5, -0.5) {};
		\node [style=point,fill=blue!10] (15) at (-1.5, -1.25) {\small$\rho$};
		\node [style=none] (16) at (1.5, 1.25) {};
		\node [style=none] (17) at (1.5, 0.5) {};
		\node [style=none] (18) at (1.5, -0.5) {};
		\node [style=none] (19) at (1.5, -.75) {};
		\node [style=none] (21) at (-1.5, -0.5) {};
		\node [style=none] (22) at (-1.5, -0.5) {};
		\node [style=upground] (23) at (1.5, 1.5) {};
		\node [style=none] (24) at (0, .75) {};
		\node [style=none] (25) at (0, -.75) {};
	\end{pgfonlayer}
	\begin{pgfonlayer}{edgelayer}
		\draw [thick gray dashed edge] (0.center) to (11.center);
		\draw [thick gray dashed edge] (11.center) to (10.center);
		\draw [thick gray dashed edge] (10.center) to (9.center);
		\draw [thick gray dashed edge] (9.center) to (8.center);
		\draw [thick gray dashed edge] (8.center) to (7.center);
		\draw [thick gray dashed edge] (7.center) to (6.center);
		\draw [thick gray dashed edge] (6.center) to (5.center);
		\draw [thick gray dashed edge] (5.center) to (4.center);
		\draw [thick gray dashed edge] (4.center) to (3.center);
		\draw [thick gray dashed edge] (3.center) to (2.center);
		\draw [thick gray dashed edge] (2.center) to (1.center);
		\draw [thick gray dashed edge] (1.center) to (0.center);
		\draw [style=qWire] (12.center) to (13.center);
		\draw [style=qWire] (14.center) to (15);
		\draw [style=qWire] (16.center) to (17.center);
		\draw [style=qWire] (18.center) to (19.center);
		\draw [qWire,bend left=90, looseness=1.50] (12.center) to (24.center);
		\draw [qWire] (24.center) to (25.center);
		\draw [qWire,bend right=90, looseness=1.50] (25.center) to (19.center);
	\end{pgfonlayer}
\end{tikzpicture}\,.
\eeq
The second process matrix is constructed similarly to before but exchanging the roles of Alice and Bob. Such a process matrix hence has a causal influence from Bob to Alice, and we denote it as $W_{\mathbf{D}\to \mathbf{A}}$. Diagrammatically,
\beq
\begin{tikzpicture}
	\begin{pgfonlayer}{nodelayer}
		\node [style=none] (0) at (-2.25, 1.5) {};
		\node [style=none] (1) at (-2.25, 1) {};
		\node [style=none] (2) at (-1.25, 1) {};
		\node [style=none] (3) at (-1.25, -1) {};
		\node [style=none] (4) at (-2.25, -1) {};
		\node [style=none] (5) at (-2.25, -1.5) {};
		\node [style=none] (6) at (2.25, -1.5) {};
		\node [style=none] (7) at (2.25, -1) {};
		\node [style=none] (8) at (1.25, -1) {};
		\node [style=none] (9) at (1.25, 1) {};
		\node [style=none] (10) at (2.25, 1) {};
		\node [style=none] (11) at (2.25, 1.5) {};
		\node [style=none] (12) at (-2, 1) {};
		\node [style=none] (13) at (-2, 0.5) {};
		\node [style=none] (14) at (-2, -0.5) {};
		\node [style=none] (15) at (-2, -1) {};
		\node [style=none] (16) at (2, 1) {};
		\node [style=none] (17) at (2, 0.5) {};
		\node [style=none] (18) at (2, -0.5) {};
		\node [style=none] (19) at (2, -1) {};
l		\node [style=none] (20) at (0, 0) {\small $W_{\mathbf{D}\to \mathbf{A}}$};
		\node [style=none] (21) at (-2, -0.5) {};
		\node [style=none] (22) at (-2, -0.5) {};
		\node [style={right label}] (23) at (-2, -0.625) {$A$};
		\node [style={right label}] (24) at (-2, 0.5) {$B$};
		\node [style={right label}] (25) at (1.375, -0.625) {$C$};
		\node [style={right label}] (26) at (1.375, 0.5) {$D$};
	\end{pgfonlayer}
	\begin{pgfonlayer}{edgelayer}
		\filldraw[fill=blue!10,draw=black] (0.center) to (11.center) to (10.center) to (9.center) to (8.center) to (7.center) to (6.center) to (5.center) to (4.center) to (3.center) to (2.center) to (1.center)to cycle;
		\draw [style=qWire] (12.center) to (13.center);
		\draw [style=qWire] (14.center) to (15.center);
		\draw [style=qWire] (16.center) to (17.center);
		\draw [style=qWire] (18.center) to (19.center);
	\end{pgfonlayer}
\end{tikzpicture}
\quad =\quad
\begin{tikzpicture}
	\begin{pgfonlayer}{nodelayer}
		\node [style=none] (0) at (2.25, 2) {};
		\node [style=none] (1) at (2.25, .75) {};
		\node [style=none] (2) at (0.75, .75) {};
		\node [style=none] (3) at (0.75, -.75) {};
		\node [style=none] (4) at (2.25, -.75) {};
		\node [style=none] (5) at (2.25, -2) {};
		\node [style=none] (6) at (-2.25, -2) {};
		\node [style=none] (7) at (-2.25, -.75) {};
		\node [style=none] (8) at (-0.75, -.75) {};
		\node [style=none] (9) at (-0.75, .75) {};
		\node [style=none] (10) at (-2.25, .75) {};
		\node [style=none] (11) at (-2.25, 2) {};
		\node [style=none] (12) at (1.5, .75) {};
		\node [style=none] (13) at (1.5, 0.5) {};
		\node [style=none] (14) at (1.5, -0.5) {};
		\node [style=point,fill=blue!10] (15) at (1.5, -1.25) {\small$\rho$};
		\node [style=none] (16) at (-1.5, 1.25) {};
		\node [style=none] (17) at (-1.5, 0.5) {};
		\node [style=none] (18) at (-1.5, -0.5) {};
		\node [style=none] (19) at (-1.5, -.75) {};
		\node [style=none] (21) at (1.5, -0.5) {};
		\node [style=none] (22) at (1.5, -0.5) {};
		\node [style=upground] (23) at (-1.5, 1.5) {};
		\node [style=none] (24) at (0, .75) {};
		\node [style=none] (25) at (0, -.75) {};
	\end{pgfonlayer}
	\begin{pgfonlayer}{edgelayer}
		\draw [thick gray dashed edge] (0.center) to (11.center);
		\draw [thick gray dashed edge] (11.center) to (10.center);
		\draw [thick gray dashed edge] (10.center) to (9.center);
		\draw [thick gray dashed edge] (9.center) to (8.center);
		\draw [thick gray dashed edge] (8.center) to (7.center);
		\draw [thick gray dashed edge] (7.center) to (6.center);
		\draw [thick gray dashed edge] (6.center) to (5.center);
		\draw [thick gray dashed edge] (5.center) to (4.center);
		\draw [thick gray dashed edge] (4.center) to (3.center);
		\draw [thick gray dashed edge] (3.center) to (2.center);
		\draw [thick gray dashed edge] (2.center) to (1.center);
		\draw [thick gray dashed edge] (1.center) to (0.center);
		\draw [style=qWire] (12.center) to (13.center);
		\draw [style=qWire] (14.center) to (15);
		\draw [style=qWire] (16.center) to (17.center);
		\draw [style=qWire] (18.center) to (19.center);
		\draw [qWire,bend right=90, looseness=1.50] (12.center) to (24.center);
		\draw [qWire](24.center) to (25.center);
		\draw [qWire,bend left=90, looseness=1.50] (25.center) to (19.center);
	\end{pgfonlayer}
\end{tikzpicture}.
\eeq

It is clear that the causal order of these two process matrices differs: in the case of $W_{\mathbf{B}\to \mathbf{C}}$, there is an influence from Alice to Bob, whilst in the case of $W_{\mathbf{D}\to \mathbf{A}}$, it goes from Bob to Alice.

Now, let us take a step back and think of the physical constituents employed to implement these process matrices, just like we did in the quantum channel discussion.
Let us denote by $\ket{\mathbf{B}\to \mathbf{C}}$ (resp., $\ket{\mathbf{D}\to \mathbf{A}}$) the state of the constituents that implement $W_{\mathbf{B}\to \mathbf{C}}$ (resp., $W_{\mathbf{D}\to \mathbf{A}}$). Then, analogously to the case of the `broken fibre', we can define a Hamiltonian which will provide us with the continuous reversible evolution of one into the other,~namely:
\beq
H = \ket{\mathbf{B}\to \mathbf{C}}\!\bra{\mathbf{D}\to \mathbf{A}} + \ket{\mathbf{D}\to \mathbf{A}}\!\bra{\mathbf{B}\to \mathbf{C}}
\eeq

We know, however, from the theorem in Ref.~\cite{castro2018dynamics} that there do not exist continuous and reversible PM-maps of the form in Equation~\eqref{eq:procMatrixTransf} which can implement this evolution. 
Hence, there are physically allowed {continuous and reversible} dynamics of a process matrix {that cannot be implemented by the PM-maps of Equation~\eqref{eq:procMatrixTransf}. Therefore, the fundamental properties of these allowed dynamics cannot be drawn from those of a PM-map described by Equation~\ref{eq:procMatrixTransf} that tries to accommodate them:  the causal order can indeed continuously and reversibly dynamically change.}

\subsection{The switch: dynamics generating indefinite causal order}\label{se:switch}

The above example shows that there exist reversible continuous dynamics which can change the causal order between two fixed definite causal orders. Can we similarly show that we can dynamically obtain indefinite causal order from definite causal order?

The answer appears to be yes, at least, if we believe that experimental implementations of the quantum switch are in principle achievable \cite{procopio2015experimental,rubino2017experimental,goswami2018indefinite}. If so, then it is simply a matter of constructing an example following the same ideas as in the previous section, which we will do next. Take one experimental realisation of the quantum switch, i.e., a setup that implements the process matrix $W_{\textsc{Switch}}$,  and denote by $\ket{\mathrm{ind}}$ the state of its physical constituents.  On the other hand, change now the state of some of the constituents, such that now the experimental setup does not implement $W_{\textsc{Switch}}$ but rather a process matrix $W_{\textsc{Def}}$ with a definite causal structure. ({{For} instance, in the case of Figure~1 of Ref.~\cite{goswami2018indefinite}, this could be achieved by having the two Polarising Beamsplitters (PBS1 and PBS2) moved out of the photon path.})
We denote this new state of the constituents by $\ket{\mathrm{def}}$.

Similarly as before, define now the Hamiltonian:
\beq\label{eq:theHam}
H = \ket{\mathrm{ind}}\!\bra{\mathrm{def}} + \ket{\mathrm{def}}\!\bra{\mathrm{ind}}\,.
\eeq
This Hamiltonian (which applies to the constituents of the experimental setup) will effectively evolve a process matrix of definite causal order ($W_{\textsc{Def}}$), described by state $\ket{\mathrm{def}}$, into one with indefinite causal order ($W_{\textsc{Switch}}$), described by state $\ket{\mathrm{ind}}$, in time $\pi/2$. As per Ref.~\cite{castro2018dynamics}, however, no continuous and reversible PM-map (Equation~\eqref{eq:procMatrixTransf}) exists that can achieve the process matrix dynamics implemented by the Hamiltonian of Equation~\eqref{eq:theHam}.
{Hence, this is another instance where the insufficiency of the framework of higher-order processes to sustain claims on the fundamental nature of the actual evolution is evidenced.}

It is worth mentioning that one can generalise the types of transformations $\mathcal{A}(t)$ of the form of Equation~\eqref{eq:theAoft} to accommodate within the framework of higher-order processes the type of example that we have presented. This is seen, for example, in Ref.~\cite{castro2018dynamics} (Sec.~VD) and Ref.~\cite{castro2020comment}. These generalisations have a crucial feature: they act on an input space larger than that of the original process matrix and output a process matrix of the same dimensionality as the one being transformed. That is, these generalisations, by definition, reduce the dimension of the input space and are therefore irreversible. Hence, the properties of the mathematical objects driving the dynamics in these generalisations do not comprehensively capture the full physical picture, and such a framework does not suffice to reflect on the fundamental nature of the actual physical evolution.

\section{Higher-order processes reexamined}\label{sec:Reexamined}

The higher-order process formalism asserts that it describes the most general way in which processes of various sorts can be transformed. Here, however, we are presenting physical examples which appear to be in contradiction with this assertion. Faced with this apparent contradiction, we must reexamine the formalism of higher-order processes {itself} in order to determine which of the assumptions going into the formalism are violated by our examples, and hence, gain an understanding of the physical situations in which using the higher-order process formalism is appropriate.

\begin{figure}
\begin{center}
\begin{tikzpicture}
	\begin{pgfonlayer}{nodelayer}
		\node [style=none] (63) at (-6, 4) {};
		\node [style=none] (64) at (-5.25, 8) {};
		\node [style=none] (65) at (3.25, 8) {};
		\node [style=none] (66) at (3.5, 4) {};
		\node [style=none] (44) at (-6.25, 6.25) {};
		\node [style=none] (46) at (-2.75, 5.25) {};
		\node [style=none] (47) at (3, 6) {};
		\node [style=none] (49) at (-1.75, 5) {};
		\node [style=none] (70) at (-5.75, 7.75) {};
		\node [style=none] (71) at (-3, 7) {};
		\node [style=none] (72) at (-3.5, 4) {};
		\node [style=none] (76) at (0.25, 3.75) {};
		\node [style=none] (77) at (-0.75, 8) {};
		\node [style=none] (78) at (2.75, 7.75) {};
		\node [style=none] (59) at (-3, -2.5) {};
		\node [style=none] (60) at (-2.25, -5.25) {};
		\node [style=none] (61) at (0, -5.5) {};
		\node [style=none] (62) at (0.75, -2.25) {};
		\node [style=small box, fill=black] (0) at (-1.75, -3.5) {$\color{white}\boldsymbol{\mathcal{A}}$};
		\node [style=none] (1) at (-1.75, -2.5) {};
		\node [style=none] (2) at (-1.75, -4.5) {};
		\node [style=small box, fill=black] (6) at (-0.5, -3.5) {$\color{white}\boldsymbol{\mathcal{B}}$};
		\node [style=none] (7) at (-0.5, -2.5) {};
		\node [style=none] (8) at (-0.5, -4.5) {};
		\node [style=small box] (12) at (-5.25, 6.75) {$A_1$};
		\node [style=none] (13) at (-5.25, 7.5) {};
		\node [style=none] (14) at (-5.25, 6) {};
		\node [style=small box] (15) at (-3.75, 5.25) {$A_2$};
		\node [style=none] (16) at (-3.75, 6) {};
		\node [style=none] (17) at (-3.75, 4.5) {};
		\node [style=small box] (18) at (2.25, 6.75) {$B_3$};
		\node [style=none] (19) at (2.25, 7.5) {};
		\node [style=none] (20) at (2.25, 6) {};
		\node [style=none] (24) at (-1, -6.25) {Black-box channels space};
		\node [style=none] (25) at (-1, 9) {Implementation Space};
		\node [style=small box] (29) at (-0.75, 5.75) {$B_1$};
		\node [style=none] (30) at (-0.75, 6.5) {};
		\node [style=none] (31) at (-0.75, 5) {};
		\node [style=small box] (32) at (0.75, 6.5) {$B_2$};
		\node [style=none] (33) at (0.75, 7.25) {};
		\node [style=none] (34) at (0.75, 5.75) {};
		\node [style=none] (45) at (-1.75, -2.25) {};
		\node [style=none] (48) at (-0.5, -2.25) {};
		\node [style=none] (53) at (5.75, 6.5) {};
		\node [style=none] (54) at (5.75, 5.5) {};
		\node [style=none] (55) at (2, -3.5) {};
		\node [style=none] (56) at (2, -4.5) {};
		\node [style=none] (57) at (4.5, -1.5) {C-Maps};
		\node [style=none] (58) at (8, 8.25) {C-Dynamics};
		\node [style=none] (67) at (-6.5, 3.5) {};
		\node [style=none] (68) at (-3.5, -2.5) {};
		\node [style=none, rotate=-65] (69) at (-6.25, 0) {Process Tomography};
	\end{pgfonlayer}
	\begin{pgfonlayer}{edgelayer}
		\draw [fill={yellow!10}] (64.center)
			 to [in=165, out=-165, looseness=1.25] (63.center)
			 to [in=-150, out=-15, looseness=0.50] (66.center)
			 to [in=-15, out=30, looseness=1.50] (65.center)
			 to [in=15, out=165, looseness=0.75] cycle;
		\draw (44.center)
			 to [in=180, out=-45, looseness=0.50] (72.center)
			 to [in=270, out=0, looseness=0.75] (46.center)
			 to [bend right=15, looseness=0.75] (71.center)
			 to [bend right=45, looseness=0.75] (70.center)
			 to [in=135, out=-150, looseness=1.25] cycle;
		\draw (49.center)
			 to [bend left] (77.center)
			 to [in=165, out=30, looseness=0.75] (78.center)
			 to [in=45, out=-15] (47.center)
			 to [in=0, out=-135, looseness=0.50] (76.center)
			 to [in=-75, out=-180] cycle;
		\draw [fill={green!10}] (60.center)
			 to [in=-135, out=150] (59.center)
			 to [in=150, out=45, looseness=0.75] (62.center)
			 to [in=15, out=-30, looseness=0.75] (61.center)
			 to [in=-30, out=-165, looseness=0.75] cycle;
		\draw [qWire](1.center) to (0);
		\draw [qWire](0) to (2.center);
		\draw [qWire](7.center) to (6);
		\draw [qWire](6) to (8.center);
		\draw [qWire](13.center) to (12);
		\draw [qWire](12) to (14.center);
		\draw [qWire](16.center) to (15);
		\draw [qWire](15) to (17.center);
		\draw [qWire](19.center) to (18);
		\draw [qWire](18) to (20.center);
		\draw [qWire](30.center) to (29);
		\draw [qWire](29) to (31.center);
		\draw [qWire](33.center) to (32);
		\draw [qWire](32) to (34.center);
		\draw [in=105, out=-45] (44.center) to (45.center);
		\draw [in=105, out=-90] (46.center) to (45.center);
		\draw [in=90, out=-135] (47.center) to (48.center);
		\draw [in=90, out=-75] (49.center) to (48.center);
		\draw [style=arrow plain, bend left=135, looseness=18.00] (53.center) to (54.center);
		\draw [style=arrow plain, bend left=135, looseness=19.25] (55.center) to (56.center);
		\draw [style=arrow plain, bend right=15] (67.center) to (68.center);
	\end{pgfonlayer}
\end{tikzpicture}
\end{center}
\caption{{{Schematic} 
 representation of where our proposed dynamics fit relative to the state of the art from the formalism of a higher-order process. In the higher-order process formalism, channels are represented by their corresponding CPTP maps---channels live in the \textit{black-box channel space} (e.g., $\boldsymbol{\mathcal{A}}$ and $\boldsymbol{\mathcal{B}}$). Dynamics within the black-box paradigm are represented by C-maps. Each such CPTP map may have different physical implementations. All these implementations live in the \textit{implementation space} (e.g., $A_1$ and $A_2$ are different implementations of $\boldsymbol{\mathcal{A}}$). Different implementations may have different evolutions, even if initially they correspond to the same CPTP map, as we discuss in this manuscript. Dynamics within the implementations space are the ones of interest in this paper. By performing process tomography of the implementations, one may recover their CPTP representation in the black-box framework.}}
\label{fig:thefig}
\end{figure}
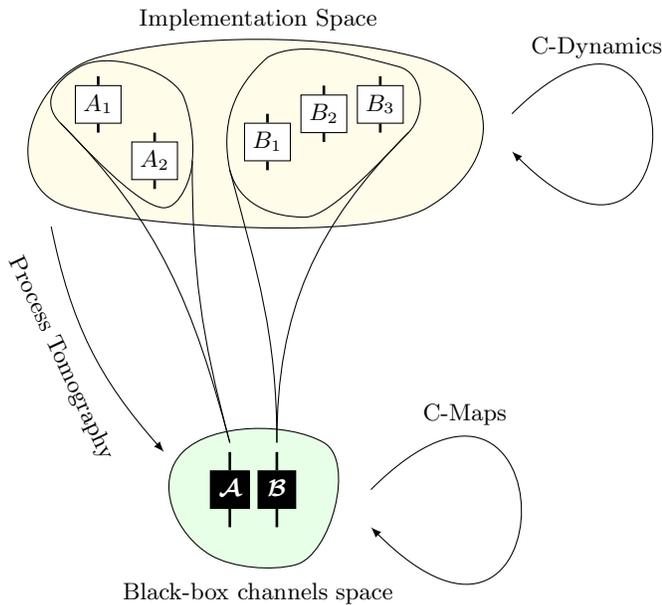

The fundamental divergence of thought {can be elucidated by} 
{considering the question ``what is a channel?''.} The higher-order process formalism views a channel simply as a CPTP map between the relevant Hilbert spaces---dynamics of channels are thus maps on the space of CPTP maps. Our examples, however, view a channel as more than just a CPTP map; that is, details of the physical implementation of the channel are also relevant---dynamics of channels are {in our examples} thus maps on the space of physical implementations of CPTP maps. This extra freedom afforded by allowing dynamics to depend not only on the CPTP map but on extra information describing the implementation is what enables a greater range of reversible and continuous evolutions {(see Figure~\ref{fig:thefig})}.
{To more clearly see this dependence on the physical implementation, consider on the one hand the experimental implementation of the switch of Figure~1 of Ref.~\cite{goswami2018indefinite} and on the other hand a modified setup whereby the beamsplitters PBS1 and PBS2 are mounted on mechanical oscillators which move the beamsplitters in and out of the photon path. The {original} setup, at least within the time-scales of the experiment, will not evolve and {hence} will for all time be associated with the {map $W_\textsc{Switch}$.} The modified set up, on the other hand, will evolve within the time-scales of the experiment and so at $t=0$ will be associated with the {map $W_\textsc{Switch}$} but at later times will be associated with some alternative $W(t)\neq W_\textsc{Switch}$ unless $t$ is some integer multiple of the time period of the oscillator. }
{We therefore see that depending on the particular implementation of $W_\textsc{Switch}$ (original vs.~modified), the process matrix $W_\textsc{Switch}$ can evolve into two different maps, and hence, its evolution cannot be fully captured by merely a higher-order process. We see clearly here how the physical implementation of the process matrix is also relevant for understanding its possible dynamics: in such cases; dynamics are maps on the space of implementations of the process matrix rather than the space of CP operators associated to the process-matrices alone.}

{A natural question hence arises:} when is it appropriate to view channels and process-matrices simply as CP maps and to describe dynamics as being dynamics on the space of such CP maps? The answer to that is {given by two conditions:} { when both (a) the} only information we have access to is the CP map associated to a channel or process-matrix, and {(b) we consider} only predictable evolution.

Condition (a) is essentially the `black-box' paradigm, in which the only information we can gain about a channel or process matrix is via process tomography over a time-scale in which the channel is assumed to be static.

Condition (b) is less commonly stated as an assumption and is most easily understood by its failure, {as we elaborate on next. Consider we are in the black-box paradigm. In this case, an evolution of a channel that depends on its implementation could still be triggered---for example, one could just poke the optical table and make the components swing around. From the point of view of the black-box paradigm, however, it would be impossible to make predictions on what the new evolution of the channel would be, since the only information we have access to is the CP map associated with the initial channel---this (unpredictable) evolution of the channel could not be represented by a higher-order process.}

{When either---or both---of these two conditions are not satisfied, which is the key for understanding} our examples, then we must view dynamics not as maps on the space of CP maps but as maps on the space of physical implementations of the channels or process-matrices. {In the next section, we propose a framework for doing just that for the case of channels.}

{If, on the other hand, {one considers only} situations in which conditions (a) and (b) are {both} satisfied, then the framework of higher-order processes is applicable. The statement of the result of {Ref.~\cite{castro2018dynamics}} may hence be more accurately expressed as follows: \blk
\begin{quotation}
\noindent \emph{{Predictable}}, reversible and continuous dynamics of \emph{{black-box}} process-matrices cannot change causal order.
\end{quotation}

 {Revisiting} 
 the question as to when causal order can evolve, we find that we can preserve the reversibility and continuity of the dynamics if we are instead willing to either abandon the black-box paradigm or predictability of the evolution---or both.}

It is worth mentioning that there are instances in which details on the physical implementation can be somehow accommodated within a black-box paradigm, as observed in Ref.~\cite{castro2018dynamics} (Sec.~VD) and Ref.~\cite{castro2020comment}, and briefly mentioned in last section. For example, in the case of process matrices, this can be accomplished by having extra input and output wires that enable one to control the state of the `environment'---in our ongoing example, this would be the position of the beamsplitters PBS1 and PBS2. Diagrammatically, this would be represented as {follows}:  
\beq\label{eq:theAt}
\begin{tikzpicture}
	\begin{pgfonlayer}{nodelayer}
		\node [style=none] (0) at (-1.75, 1.5) {};
		\node [style=none] (1) at (-1.75, 1) {};
		\node [style=none] (2) at (-0.75, 1) {};
		\node [style=none] (3) at (-0.75, -1) {};
		\node [style=none] (4) at (-1.75, -1) {};
		\node [style=none] (5) at (-1.75, -1.5) {};
		\node [style=none] (6) at (1.75, -1.5) {};
		\node [style=none] (7) at (1.75, -1) {};
		\node [style=none] (8) at (0.75, -1) {};
		\node [style=none] (9) at (0.75, 1) {};
		\node [style=none] (10) at (1.75, 1) {};
		\node [style=none] (11) at (1.75, 1.5) {};
		\node [style=none] (12) at (-1.5, 1) {};
		\node [style=none] (13) at (-1.5, 0.5) {};
		\node [style=none] (14) at (-1.5, -0.5) {};
		\node [style=none] (15) at (-1.5, -1) {};
		\node [style=none] (16) at (1.5, 1) {};
		\node [style=none] (17) at (1.5, 0.5) {};
		\node [style=none] (18) at (1.5, -0.5) {};
		\node [style=none] (19) at (1.5, -1) {};
		\node [style=none] (20) at (0, 0) {\small $W(t)$};
		\node [style=none] (21) at (-1.5, -0.5) {};
		\node [style=none] (22) at (-1.5, -0.5) {};
		\node [style=right label] (23) at (-1.5, -0.675) {$A$};
		\node [style=right label] (24) at (-1.5, 0.5) {$B$};
		\node [style=right label] (25) at (.875, -0.675) {$C$};
		\node [style=right label] (26) at (.875, 0.5) {$D$};
	\end{pgfonlayer}
	\begin{pgfonlayer}{edgelayer}
			\filldraw[fill=blue!10,draw=black] (0.center) to (11.center) to (10.center) to (9.center) to (8.center) to (7.center) to (6.center) to (5.center) to (4.center) to (3.center) to (2.center) to (1.center)to cycle;
		\draw (0.center) to (11.center);
		\draw (11.center) to (10.center);
		\draw (10.center) to (9.center);
		\draw (9.center) to (8.center);
		\draw (8.center) to (7.center);
		\draw (7.center) to (6.center);
		\draw (6.center) to (5.center);
		\draw (5.center) to (4.center);
		\draw (4.center) to (3.center);
		\draw (3.center) to (2.center);
		\draw (2.center) to (1.center);
		\draw (1.center) to (0.center);
		\draw [style=qWire] (12.center) to (13.center);
		\draw [style=qWire] (14.center) to (15.center);
		\draw [style=qWire] (16.center) to (17.center);
		\draw [style=qWire] (18.center) to (19.center);
	\end{pgfonlayer}
\end{tikzpicture}
\quad = \quad
\begin{tikzpicture}
	\begin{pgfonlayer}{nodelayer}
		\node [style=none] (12) at (-1.5, 1) {};
		\node [style=none] (13) at (-1.5, 0.5) {};
		\node [style=none] (14) at (-1.5, -0.5) {};
		\node [style=none] (15) at (-1.5, -1) {};
		\node [style=none] (16) at (1.5, 1) {};
		\node [style=none] (17) at (1.5, 0.5) {};
		\node [style=none] (18) at (1.5, -0.5) {};
		\node [style=none] (19) at (1.5, -1) {};
		\node [style=none] (21) at (-1.5, -0.5) {};
		\node [style=none] (22) at (-1.5, -0.5) {};
		\node [style=right label] (23) at (-1.5, -0.875) {$A$};
		\node [style=right label] (24) at (-1.5, 0.675) {$B$};
		\node [style=right label] (25) at (0.75, -0.875) {$C$};
		\node [style=right label] (26) at (0.75, 0.675) {$D$};
		\node [style=none] (27) at (-1, -0.5) {};
		\node [style=none] (28) at (-1, 0.5) {};
		\node [style=none] (29) at (-2.25, 0.5) {};
		\node [style=none] (30) at (-2.25, -0.5) {};
		\node [style=none] (31) at (-2.25, -2) {};
		\node [style=none] (32) at (2.25, -2) {};
		\node [style=none] (33) at (2.25, -0.5) {};
		\node [style=none] (34) at (1, -0.5) {};
		\node [style=none] (35) at (1, 0.5) {};
		\node [style=none] (36) at (2.25, 0.5) {};
		\node [style=none] (37) at (2.25, 2) {};
		\node [style=none] (38) at (-2.25, 2) {};
		\node [style=none] (39) at (-4, 3) {};
		\node [style=none] (40) at (-4, 1) {};
		\node [style=none] (41) at (-3, 1) {};
		\node [style=none] (42) at (-3, -1) {};
		\node [style=none] (43) at (-4, -1) {};
		\node [style=none] (44) at (-4, -2.75) {};
		\node [style=none] (45) at (4, -2.75) {};
		\node [style=none] (46) at (4, -1) {};
		\node [style=none] (47) at (3, -1) {};
		\node [style=none] (48) at (3, 1) {};
		\node [style=none] (49) at (4, 1) {};
		\node [style=none] (50) at (4, 3) {};
		\node [style=none] (51) at (-3.75, 1) {};
		\node [style=none] (52) at (-3.75, 0.5) {};
		\node [style=none] (53) at (-3.75, -0.5) {};
		\node [style=none] (54) at (-3.75, -1) {};
		\node [style=none] (55) at (3.75, 1) {};
		\node [style=none] (56) at (3.75, 0.5) {};
		\node [style=none] (57) at (3.75, -0.5) {};
		\node [style=none] (58) at (3.75, -1) {};
		\node [style=none] (59) at (-3.75, -0.5) {};
		\node [style=none] (60) at (-3.75, -0.5) {};
		\node [style=right label] (61) at (-3.75, -0.675) {$A$};
		\node [style=right label] (62) at (-3.75, 0.5) {$B$};
		\node [style=right label] (63) at (3.125, -0.675) {$C$};
		\node [style=right label] (64) at (3.125, 0.5) {$D$};
		\node [style=none] (65) at (0, 2.5) {\small $\widetilde{\mathcal{A}}(t)$};
		\node [style=none] (66) at (-1.75, 1) {};
		\node [style=none] (67) at (-1.75, 1.5) {};
		\node [style=none] (68) at (1.75, 1.5) {};
		\node [style=none] (69) at (1.75, 1) {};
		\node [style=none] (70) at (0.75, 1) {};
		\node [style=none] (71) at (-0.75, 1) {};
		\node [style=none] (72) at (-0.75, -1) {};
		\node [style=none] (73) at (-2, -1) {};
		\node [style=none] (74) at (-2, -1.5) {};
		\node [style=none] (75) at (2, -1.5) {};
		\node [style=none] (76) at (2, -1) {};
		\node [style=none] (77) at (0.75, -1) {};
		\node [style=none] (78) at (0, 0) {\small $\widetilde{W}$};
		\node [style=none] (79) at (0, -1.5) {};
		\node [style=none] (80) at (0, -2) {};
		\node [style=none] (81) at (0, 2) {};
		\node [style=none] (82) at (0, 1.5) {};
		\node [style=right label] (83) at (0, -1.875) {$E_i$};
		\node [style=right label] (84) at (0, 1.625) {$E_o$};
	\end{pgfonlayer}
	\begin{pgfonlayer}{edgelayer}
		\draw [fill={purple!10}, draw=black] (39.center)
			 to (50.center)
			 to (49.center)
			 to (48.center)
			 to (47.center)
			 to (46.center)
			 to (45.center)
			 to (44.center)
			 to (43.center)
			 to (42.center)
			 to (41.center)
			 to (40.center)
			 to cycle;
		\draw [fill=white, draw=black] (38.center)
			 to (37.center)
			 to (36.center)
			 to (35.center)
			 to (34.center)
			 to (33.center)
			 to (32.center)
			 to (31.center)
			 to (30.center)
			 to (27.center)
			 to (28.center)
			 to (29.center)
			 to cycle;
		\draw [style=qWire] (12.center) to (13.center);
		\draw [style=qWire] (14.center) to (15.center);
		\draw [style=qWire] (16.center) to (17.center);
		\draw [style=qWire] (18.center) to (19.center);
		\draw [style=qWire] (51.center) to (52.center);
		\draw [style=qWire] (53.center) to (54.center);
		\draw [style=qWire] (55.center) to (56.center);
		\draw [style=qWire] (57.center) to (58.center);
		\draw [fill={blue!10}, draw=black] (67.center)
			 to (68.center)
			 to (69.center)
			 to (70.center)
			 to (77.center)
			 to (76.center)
			 to (75.center)
			 to (74.center)
			 to (73.center)
			 to (72.center)
			 to (71.center)
			 to (66.center)
			 to cycle;
		\draw (80.center) to (79.center);
		\draw (82.center) to (81.center);
	\end{pgfonlayer}
\end{tikzpicture}
\,,
\eeq
where $\mathcal{\widetilde{A}}(t)$ drives the evolution of a process $\widetilde{W}$, such that $\mathcal{\widetilde{A}}(0)[\widetilde{W}] = W_\textsc{Switch}$. Now, since $\mathcal{\widetilde{A}}(t)[\widetilde{W}]$ should have no ``open wires'' (which ensures that statements about the causal properties of the evolved process matrix can be made in this paradigm), then by definition $\mathcal{\widetilde{A}}(t)$ must be an irreversible transformation \cite{castro2018dynamics,castro2020comment}. However, this irreversibility is not a signature of fundamentally irreversible dynamics but merely a result of the particular paradigm within which the situation is being expressed. This mathematical feature was mentioned by Refs.~\cite{castro2018dynamics,castro2020comment}, who make the assumption of ``no open wires'' for their main result, although they do not elaborate on the physical relevance of the applicability of such an assumption. 
In the next section, we will present an alternative description to channels dynamics tailored at distilling the properties of the physical dynamics at its fundamental~level.

\section{Alternative description of channel dynamics}\label{app:alternative}

The failure of the paradigm of higher-order processes to 
{reveal the fundamental properties of the laws of nature driving the physical dynamics of channels and process matrices  is highlighted by the examples presented in the previous sections.}
Ultimately, however, to know what dynamics are in principle achievable would require us to have a complete microscopic description of the constituents of the object (i.e., channel or process matrix), how these constituents can evolve in time, and, moreover, how they interact with the system of interest.

This is far too fine-grained a perspective to be of practical use. Therefore, we need to consider whether we actually need this full fine-grained picture {or whether there is an alternative paradigm to higher-order processes which we can use to describe the physical scenarios that we have discussed {in this paper} without having to {resort} to the full microscopic description.}

In this section, we propose {such} an alternate paradigm for describing the evolution of channels. Here, we {characterise}
a channel $\mathcal{E}$
as a description of the interaction $\mathcal{X}$ between the input system $A$ and the physical constituents of the channel, which are described by some physical system $C$. Diagrammatically, this is {expressed as}
\beq \label{eq:Stinespring}
\begin{tikzpicture}
	\begin{pgfonlayer}{nodelayer}
		\node [style={small box},fill=blue!10] (0) at (0, -0) {\small $\mathcal{E}$};
		\node [style=none] (1) at (0, 1) {};
		\node [style=none] (2) at (0, -1) {};
		\node [style=right label] (3) at (0, -1) {$A$};
		\node [style=right label] (4) at (0, 1) {$B$};
	\end{pgfonlayer}
	\begin{pgfonlayer}{edgelayer}
		\draw [qWire] (2.center) to (0);
		\draw [qWire] (0) to (1.center);
	\end{pgfonlayer}
\end{tikzpicture}
\quad =\quad
\begin{tikzpicture}
	\begin{pgfonlayer}{nodelayer}
		\node [style=point,fill=blue!10] (0) at (-0.7500001, -1.75) {\small$\phi$};
		\node [style=none] (1) at (-0.7500001, -0.5000001) {};
		\node [style={right label}] (2) at (-0.7500001, -1) {$C$};
		\node [style=none] (3) at (0.7500001, -0.5000001) {};
		\node [style=none] (4) at (0.7500001, -3) {};
		\node [style={right label}] (5) at (0.7500001, -3) {$A$};
		\node [style={right label}] (6) at (0.7500001, 2) {$B$};
		\node [style=none] (7) at (0.7500001, 0.5000001) {};
		\node [style=none] (8) at (0.7500001, 2) {};
		\node [style=none] (9) at (0, -0) {\small $\mathcal{X}$};
		\node [style=none] (10) at (1.25, 0.5000001) {};
		\node [style=none] (11) at (1.25, -0.5000001) {};
		\node [style=none] (12) at (-1.25, -0.5000001) {};
		\node [style=none] (13) at (-1.25, 0.5000001) {};
		\node [style=none] (14) at (-0.7500001, 0.5000001) {};
		\node [style=none] (15) at (-0.7500001, 1.25) {};
		\node [style={right label}] (16) at (-0.7500001, 0.7500001) {$C'$};
		\node [style=upground] (17) at (-0.7500001, 1.5) {};
	\end{pgfonlayer}
	\begin{pgfonlayer}{edgelayer}
		\draw [qWire] (1.center) to (0);
		\draw [qWire] (3.center) to (4.center);
		\draw [qWire] (7.center) to (8.center);
		\filldraw[draw=black,fill=blue!10] (13.center) to (10.center) to (11.center) to (12.center) to cycle;
		\draw[style=qWire] (15.center) to (14.center);
	\end{pgfonlayer}
\end{tikzpicture} \,.
\eeq
That is, we have the system $C$ representing the initial internal degrees of freedom of the channel, which starts in some state $\phi$; then, there is some microscopic interaction between these degrees of freedom and the system of interest $A$. On a fundamental level, we expect the interaction to be reversible, that is, to be some unitary transformation. Therefore, it should be the case that $C\otimes A \cong C'\otimes B$. In the end, we are only interested in the effective evolution of the system $A$ into the system $B$, and so we trace out system $C'$.

Now, from this perspective, {it can readily be seen} how this channel should evolve in time. The channel dynamics will simply be described by the evolution of the physical constituents of the channel, i.e., the evolution in time of the {state $\phi$}:
\beq
\begin{tikzpicture}
	\begin{pgfonlayer}{nodelayer}
		\node [style={small box},fill=blue!10] (0) at (0, -0) {\small $\mathcal{E}(t)$};
		\node [style=none] (1) at (0, 1) {};
		\node [style=none] (2) at (0, -1) {};
		\node [style=right label] (3) at (0, -1) {$A$};
		\node [style=right label] (4) at (0, 1) {$B$};
	\end{pgfonlayer}
	\begin{pgfonlayer}{edgelayer}
		\draw [qWire] (2.center) to (0);
		\draw [qWire] (0) to (1.center);
	\end{pgfonlayer}
\end{tikzpicture}
\quad =\quad
\begin{tikzpicture}
	\begin{pgfonlayer}{nodelayer}
		\node [style=point,fill=blue!10] (0) at (-0.7500001, -1.75) {\small $\phi(t)$};
		\node [style=none] (1) at (-0.7500001, -0.5000001) {};
		\node [style={right label}] (2) at (-0.7500001, -1) {$C$};
		\node [style=none] (3) at (0.7500001, -0.5000001) {};
		\node [style=none] (4) at (0.7500001, -3) {};
		\node [style={right label}] (5) at (0.7500001, -3) {$A$};
		\node [style={right label}] (6) at (0.7500001, 2) {$B$};
		\node [style=none] (7) at (0.7500001, 0.5000001) {};
		\node [style=none] (8) at (0.7500001, 2) {};
		\node [style=none] (9) at (0, -0) {\small $\mathcal{X}$};
		\node [style=none] (10) at (1.25, 0.5000001) {};
		\node [style=none] (11) at (1.25, -0.5000001) {};
		\node [style=none] (12) at (-1.25, -0.5000001) {};
		\node [style=none] (13) at (-1.25, 0.5000001) {};
		\node [style=none] (14) at (-0.7500001, 0.5000001) {};
		\node [style=none] (15) at (-0.7500001, 1.25) {};
		\node [style={right label}] (16) at (-0.7500001, 0.7500001) {$C'$};
		\node [style=upground] (17) at (-0.7500001, 1.5) {};
	\end{pgfonlayer}
	\begin{pgfonlayer}{edgelayer}
		\draw [qWire] (1.center) to (0);
		\draw [qWire] (3.center) to (4.center);
		\draw [qWire] (7.center) to (8.center);
		\filldraw[draw=black,fill=blue!10] (13.center) to (10.center) to (11.center) to (12.center) to cycle;
		\draw[style=qWire] (15.center) to (14.center);
	\end{pgfonlayer}
\end{tikzpicture}\,.
\eeq
The continuous reversible evolution would then simply correspond to the continuous reversible evolution of the state of the system $C$, that is, when we can further decompose this {description as}
\beq\label{eq:ourev}
\begin{tikzpicture}
	\begin{pgfonlayer}{nodelayer}
		\node [style={small box},fill=blue!10] (0) at (0, -0) {\small $\mathcal{E}(t)$};
		\node [style=none] (1) at (0, 1) {};
		\node [style=none] (2) at (0, -1) {};
		\node [style=right label] (3) at (0, -1) {$A$};
		\node [style=right label] (4) at (0, 1) {$B$};
	\end{pgfonlayer}
	\begin{pgfonlayer}{edgelayer}
		\draw [qWire] (2.center) to (0);
		\draw [qWire] (0) to (1.center);
	\end{pgfonlayer}
\end{tikzpicture}
\quad =\quad
\begin{tikzpicture}
	\begin{pgfonlayer}{nodelayer}
		\node [style=point,fill=blue!10] (0) at (-0.7500001, -3.5) {\small $\phi(0)$};
		\node [style={small box},fill=purple!10] (1) at (-0.7500001, -1.75) {\small $e^{iHt}$};
		\node [style={right label}] (2) at (-0.7500001, -2.675) {$C$};
		\node [style=none] (3) at (0.7500001, -0.5000001) {};
		\node [style=none] (4) at (0.7500001, -5) {};
		\node [style={right label}] (5) at (0.7500001, -5) {$A$};
		\node [style={right label}] (6) at (0.7500001, 2) {$B$};
		\node [style=none] (7) at (0.7500001, 0.5000001) {};
		\node [style=none] (8) at (0.7500001, 2) {};
		\node [style=none] (9) at (0, -0) {\small $\mathcal{X}$};
		\node [style=none] (10) at (1.25, 0.5000001) {};
		\node [style=none] (11) at (1.25, -0.5000001) {};
		\node [style=none] (12) at (-1.25, -0.5000001) {};
		\node [style=none] (13) at (-1.25, 0.5000001) {};
		\node [style=none] (14) at (-0.7500001, 0.5000001) {};
		\node [style=none] (15) at (-0.7500001, 1.25) {};
		\node [style={right label}] (16) at (-0.7500001, 0.7500001) {$C'$};
		\node [style=upground] (17) at (-0.7500001, 1.5) {};
		\node [style=none] (18) at (-0.7500001, -0.5000001) {};
		\node [style={right label}] (19) at (-0.7500001, -1) {$C$};
	\end{pgfonlayer}
	\begin{pgfonlayer}{edgelayer}
		\draw [qWire] (1) to (0);
		\draw [qWire] (3.center) to (4.center);
		\draw [qWire] (7.center) to (8.center);
		\filldraw[draw=black,fill=blue!10] (13.center) to (10.center) to (11.center) to (12.center) to cycle;
		\draw [style=qWire] (15.center) to (14.center);
		\draw [qWire] (18.center) to (1);
	\end{pgfonlayer}
\end{tikzpicture}\,,
\eeq
where $H$ is the relevant Hamiltonian for the internal degrees of freedom of the channel.

In general, as mentioned above, we will not know the precise microscopic description of the channel and its interaction with the system of interest. We can, however, notice that Equation~\ref{eq:Stinespring} is nothing but a Stinespring dilation \cite{dilation} of the channel $\mathcal{E}$. We therefore propose that if the only information we have available to us is the description of a channel as a CPTP map, then {the  way} to assess the dynamics is to look at the dynamics possible for arbitrary dilations of the channel.

It is worth at this stage highlighting the similarities and differences to the approach of Refs.~\cite{castro2018dynamics,castro2020comment} summarised in the last section. On the one hand, one can interpret the components of the right-hand side of Equation~\eqref{eq:ourev} as a particular form of the higher-order process $\mathcal{\widetilde{A}}(t)$ of Equation~\eqref{eq:theAt} for the case of channels {evolution}:
\beq
\begin{tikzpicture}
	\begin{pgfonlayer}{nodelayer}
		\node [style=none] (0) at (-1, 0.5) {};
		\node [style=none] (1) at (-1, 1.25) {};
		\node [style=none] (2) at (-1, -1.25) {};
		\node [style=right label] (3) at (-1, -1) {$C$};
		\node [style=right label] (4) at (-1, 0.75) {$C'$};
		\node [style=none] (5) at (-1.5, 1.25) {};
		\node [style=none] (6) at (-1.5, 1.75) {};
		\node [style=none] (7) at (1, 1.25) {};
		\node [style=none] (8) at (1, -1.25) {};
		\node [style=none] (9) at (-1.5, -1.25) {};
		\node [style=none] (10) at (-1.5, -1.75) {};
		\node [style=none] (11) at (2.5, -1.75) {};
		\node [style=none] (12) at (2.5, 1.75) {};
		\node [style=none] (13) at (1.75, 0) {\small $\widetilde{\mathcal{A}}(t)$};
		\node [style=right label] (14) at (1.75, -2.5) {$A$};
		\node [style=none] (15) at (1.75, -1.75) {};
		\node [style=none] (16) at (1.75, -2.5) {};
		\node [style=none] (17) at (1.75, 2.5) {};
		\node [style=right label] (18) at (1.75, 2.5) {$B$};
		\node [style=none] (19) at (1.75, 1.75) {};
		\node [style=none] (20) at (-1, -0.5) {};
		\node [style=none] (21) at (0, 0.5) {};
		\node [style=none] (22) at (0, 1.25) {};
		\node [style=none] (23) at (0, -1.25) {};
		\node [style=right label] (24) at (0, -1) {$A$};
		\node [style=right label] (25) at (0, 0.75) {$B$};
		\node [style=none] (26) at (0, -0.5) {};
	\end{pgfonlayer}
	\begin{pgfonlayer}{edgelayer}
		\draw [qWire] (2.center) to (20.center);
		\draw [qWire] (0.center) to (1.center);
		\draw [fill={purple!10}, draw=black] (5.center)
			 to (7.center)
			 to (8.center)
			 to (9.center)
			 to (10.center)
			 to (11.center)
			 to (12.center)
			 to (6.center)
			 to cycle;
		\draw [qWire] (16.center) to (15.center);
		\draw [qWire] (19.center) to (17.center);
		\draw [qWire] (23.center) to (26.center);
		\draw [qWire] (21.center) to (22.center);
	\end{pgfonlayer}
\end{tikzpicture}
\ \ = \ \ 
\begin{tikzpicture}
	\begin{pgfonlayer}{nodelayer}
		\node [style=point, fill={blue!10}] (0) at (-0.5, -1.75) {\small $\phi(t)$};
		\node [style=none] (1) at (-0.5, -0.5) {};
		\node [style=right label] (2) at (-0.5, -0.75) {$C$};
		\node [style=none] (3) at (0.75, -0.5) {};
		\node [style=none] (4) at (0.75, -1) {};
		\node [style=right label] (5) at (2, -3.5) {$A$};
		\node [style=right label] (6) at (2, 2.5) {$B$};
		\node [style=none] (7) at (0.75, 0.5) {};
		\node [style=none] (8) at (0.75, 1) {};
		\node [style=none] (14) at (-0.5, 0.5) {};
		\node [style=none] (15) at (-0.5, 1.25) {};
		\node [style=right label] (16) at (-0.5, 0.5) {$C'$};
		\node [style=upground] (17) at (-0.5, 1.5) {};
		\node [style=none] (18) at (1.5, 1) {};
		\node [style=none] (19) at (1.5, -1) {};
		\node [style=none] (20) at (-1.75, -1) {};
		\node [style=none] (21) at (-1.75, -3) {};
		\node [style=none] (22) at (2.5, -3) {};
		\node [style=none] (23) at (2.5, 2.25) {};
		\node [style=none] (24) at (-1.75, 2.25) {};
		\node [style=none] (25) at (-1.75, 1) {};
		\node [style=right label] (26) at (0.75, -0.75) {$A$};
		\node [style=right label] (27) at (0.75, 0.5) {$B$};
		\node [style=none] (28) at (2, 2.25) {};
		\node [style=none] (29) at (2, 2.75) {};
		\node [style=none] (30) at (0.75, 1) {};
		\node [style=none] (31) at (2, 2.25) {};
		\node [style=none] (32) at (2, -3.5) {};
		\node [style=none] (33) at (2, -3) {};
		\node [style=none] (34) at (2, -3) {};
		\node [style=none] (35) at (0.75, -1) {};
	\end{pgfonlayer}
	\begin{pgfonlayer}{edgelayer}
		\draw [qWire] (1.center) to (0);
		\draw [qWire] (3.center) to (4.center);
		\draw [qWire] (7.center) to (8.center);
		\draw [style=qWire] (15.center) to (14.center);
		\draw [thick gray dashed edge] (23.center)
			 to (24.center)
			 to (25.center)
			 to (18.center)
			 to (19.center)
			 to (20.center)
			 to (21.center)
			 to (22.center)
			 to cycle;
		\draw [qWire] (28.center) to (29.center);
		\draw [qWire, in=270, out=90] (30.center) to (31.center);
		\draw [qWire] (32.center) to (33.center);
		\draw [qWire, in=270, out=90] (34.center) to (35.center);
	\end{pgfonlayer}
\end{tikzpicture}
\,,
\eeq
which acts on the process $\chi$.
However, this particular form that we take allows us to assess the nature of the dynamics based on the properties of the evolution $\phi(t)$ rather than on the properties of the irreversible process $\mathcal{\widetilde{A}}(t)$. ({{{Notice} that this generic irreversibility stems from the caveat that system $C'$ is being traced out, since the partial trace of an overall reversible process results in an irreversible one.}) 
That is how we can correctly identify as `continuous and reversible' certain dynamics that do not look quite so from the point of view of $\mathcal{\widetilde{A}}(t)$.}

{To see the scope of application of our framework, let us finally present a concrete example. Consider} the following dilation of the identity {channel}:
\beq
\begin{tikzpicture}
	\begin{pgfonlayer}{nodelayer}
		\node [style=none] (0) at (0, 2.25) {};
		\node [style=none] (1) at (0, -4) {};
	\end{pgfonlayer}
	\begin{pgfonlayer}{edgelayer}
		\draw [qWire](0.center) to (1.center);
	\end{pgfonlayer}
\end{tikzpicture}
\quad = \quad
\begin{tikzpicture}
	\begin{pgfonlayer}{nodelayer}
		\node [style=none] (0) at (0.25, 0) {$\textsc{Swap}$};
		\node [style=none] (1) at (-1, 0.5) {};
		\node [style=none] (2) at (-1, -0.5) {};
		\node [style=none] (3) at (1.5, -0.5) {};
		\node [style=none] (4) at (1.5, 0.5) {};
		\node [style=black dot] (5) at (-4, 0) {};
		\node [style=none] (6) at (-1, 0) {};
		\node [style=none] (7) at (-0.5, 0.5) {};
		\node [style=none] (8) at (-0.5, 1.5) {};
		\node [style=none] (9) at (1, 0.5) {};
		\node [style=none] (10) at (1, 2.25) {};
		\node [style=none] (11) at (-0.5, -2.25) {};
		\node [style=none] (12) at (-0.5, -0.5) {};
		\node [style=none] (13) at (1, -4) {};
		\node [style=none] (14) at (1, -0.5) {};
		\node [style=none] (15) at (-4, -2.75) {};
		\node [style=point,fill=blue!10] (16) at (-4, -2.625) {\small $0$};
		\node [style=none] (17) at (-1.25, -2.75) {\small $\Psi^+$};
		\node [style=none] (18) at (-3, -2.25) {};
		\node [style=none] (19) at (0.25, -2.25) {};
		\node [style=none] (20) at (-1.25, -3.5) {};
		\node [style=none] (21) at (-2.25, 1.5) {};
		\node [style=none] (22) at (-4, 1.5) {};
		\node [style=none] (23) at (-2.25, -2.25) {};
		\node [style=upground] (24) at (-4, 1.75) {};
		\node [style=upground] (25) at (-2.25, 1.75) {};
		\node [style=upground] (26) at (-0.5, 1.75) {};
	\end{pgfonlayer}
	\begin{pgfonlayer}{edgelayer}
		\filldraw[draw=black,fill=blue!10] (1.center) to (4.center) to (3.center) to (2.center) to cycle;
		\draw (5) to (6.center);
		\draw [qWire] (8.center) to (7.center);
		\draw [qWire] (10.center) to (9.center);
		\draw [qWire] (12.center) to (11.center);
		\draw [qWire] (14.center) to (13.center);
		\draw [qWire] (5) to (15.center);
		\filldraw[draw=black,fill=blue!10] (18.center) to (20.center) to (19.center) to cycle;
		\draw [qWire] (22.center) to (5);
		\draw [qWire] (21.center) to (23.center);
	\end{pgfonlayer}
\end{tikzpicture}\,.
\eeq
Then, under the Hamiltonian evolution $H=X\otimes I\otimes I$ for the dilating system, we {obtain}
\beq
\begin{tikzpicture}
	\begin{pgfonlayer}{nodelayer}
		\node [style={small box},fill=blue!10] (0) at (0, -0) {\small $\mathcal{E}(t)$};
		\node [style=none] (1) at (0, 2) {};
		\node [style=none] (2) at (0, -2) {};
	\end{pgfonlayer}
	\begin{pgfonlayer}{edgelayer}
		\draw [qWire] (2.center) to (0);
		\draw [qWire] (0) to (1.center);
	\end{pgfonlayer}
\end{tikzpicture}
\quad =\quad
\begin{tikzpicture}
	\begin{pgfonlayer}{nodelayer}
		\node [style=none] (0) at (0.25, 0) {$\textsc{Swap}$};
		\node [style=none] (1) at (-1, 0.5) {};
		\node [style=none] (2) at (-1, -0.5) {};
		\node [style=none] (3) at (1.5, -0.5) {};
		\node [style=none] (4) at (1.5, 0.5) {};
		\node [style=black dot] (5) at (-4, 0) {};
		\node [style=none] (6) at (-1, 0) {};
		\node [style=none] (7) at (-0.5, 0.5) {};
		\node [style=none] (8) at (-0.5, 1.5) {};
		\node [style=none] (9) at (1, 0.5) {};
		\node [style=none] (10) at (1, 2.25) {};
		\node [style=none] (11) at (-0.5, -2.25) {};
		\node [style=none] (12) at (-0.5, -0.5) {};
		\node [style=none] (13) at (1, -4) {};
		\node [style=none] (14) at (1, -0.5) {};
		\node [style=small box,fill=purple!10] (15) at (-4, -1.325) {\small $e^{iXt}$};
		\node [style=point,fill=blue!10] (16) at (-4, -2.625) {\small $0$};
		\node [style=none] (17) at (-1.25, -2.75) {\small $\Psi^+$};
		\node [style=none] (18) at (-3, -2.25) {};
		\node [style=none] (19) at (0.25, -2.25) {};
		\node [style=none] (20) at (-1.25, -3.5) {};
		\node [style=none] (21) at (-2.25, 1.5) {};
		\node [style=none] (22) at (-4, 1.5) {};
		\node [style=none] (23) at (-2.25, -2.25) {};
		\node [style=upground] (24) at (-4, 1.75) {};
		\node [style=upground] (25) at (-2.25, 1.75) {};
		\node [style=upground] (26) at (-0.5, 1.75) {};
	\end{pgfonlayer}
	\begin{pgfonlayer}{edgelayer}
		\filldraw[draw=black,fill=blue!10] (1.center) to (4.center) to (3.center) to (2.center) to cycle;
		\draw (5) to (6.center);
		\draw [qWire] (8.center) to (7.center);
		\draw [qWire] (10.center) to (9.center);
		\draw [qWire] (12.center) to (11.center);
		\draw [qWire] (14.center) to (13.center);
		\draw [qWire] (5) to (15);
		\draw [qWire] (15) to (16);
		\filldraw[draw=black,fill=blue!10] (18.center) to (20.center) to (19.center) to cycle;
		\draw [qWire] (22.center) to (5);
		\draw [qWire] (21.center) to (23.center);
	\end{pgfonlayer}
\end{tikzpicture}\,.
\eeq
Under these dynamics, $\mathcal{E}(0)$ is the identity channel we started with, whilst, by $t=\pi/2$, the channel $\mathcal{E}(t)$ will have evolved to the depolarising channel as per our first example. We therefore see that this view of dynamics arising from channel dilations is {more comprehensive than the frameworks arising from higher-order transformations, as it captures the underlying physical nature of other relevant physical situations.}

One crucial assumption is made here that we should make explicit. Viewing the physical constituents of the channel as a sort of environment for the `system which is to be transformed by the channel', we assume the following: (i) the state of this environment $C$ changes in time, while (ii) the coupling dynamics between $A$ and $C$  stay time
independent. For any practical purposes, this assumption means that the time-scale of the coupling $\mathcal{X}$ is much shorter than  the time-scale of the
environment evolution; hence, the environment state seems frozen from the perspective of the channel and its input state.
This assumption holds, for instance, in natural physical situations like the recently developed schemes of system-entanglement detection \cite{Cywinski_et_al_2019},
when the environment Hamiltonian has a small norm if compared to the coupling constant.
The assumption will no longer hold, however, in the less natural scenario where the roles of the environment and the systems are swapped. Indeed, the assumed time-scales of the coupling and system dynamics (now playing the role of the environment) are just opposite to those required in our scenario.

\section{Conclusion}

{The study of the fundamental dynamics of process matrices was first raised in Ref.~\cite{castro2018dynamics}. There, a particular question was that of whether or not the causal order set by a process matrix would be preserved by continuous and reversible dynamics.
This problem plays a fundamental role in the study of the information processing power that the resources allowed by quantum (and even post-quantum) theories enable \cite{qswitch,araujo2014computational,feix2015quantum,guerin2016exponential,kristjansson2020resource,zhao2020quantum}.} 
 Ref.~\cite{castro2018dynamics} takes a black-box approach to this problem and draws conclusions regarding the possibility or impossibility of the dynamical evolution of quantum devices based on the information given by the black-box description of such devices (i.e., the CPTP map associated to a quantum channel, rather than the physical description of the components that implement that channel in the lab). 
In this work, we revisit the question of which possible dynamics we can expect objects, such as channels and process matrices, to feature.
We found that the commonly used formalism of higher-order processes  (that is, the black-box approach endorsed by Ref.~\cite{castro2018dynamics})  may mislead our grasp on the fundamentals of physical dynamics, and we presented examples of continuous and reversible dynamics that may change the causal order of the processes they are applied to. This is of particular interest, since continuous and reversible transformations are in the core of some axiomatic reconstructions of Quantum Theory \cite{hardy2001quantum,dakic2011quantum,masanes2011derivation}, and one hence needs to understand their full scope. Our arguments stem from situations where one can apply Hamiltonian dynamics to the description of the physical constituents that implement the corresponding processes whose evolution is being studied. We acknowledge that such a  description is not always possible to provide either at a fundamental level (e.g., when one studies physical theories beyond quantum theory or from the perspective of Operational Probabilistic Theories) or a practical level, where sometimes higher-order processes provide our best understanding of the situation. \blk  

The examples we discussed range from simple mathematical setups to more complex physical ones. In particular, the explicit time dependence of the transformations we found raises the natural question of how to understand superchannels with a quantum memory in continuous time, which we defer to future research.
{Another interesting question is whether the dynamics of the objects (be they channels or process matrices) could be driven by a different kind of physical mechanism than the ones we discuss here. In particular, could quantum scattering theory be leveraged to change the causal structure of a process? Whether or not this is possible is left for future research.}

{In this work, we moreover} propose a more refined formalism for studying the {{reversible} continuous time} dynamics of channels, which subsumes {and extends} {the dynamics of channels via higher-order processes.} The main idea behind our construction comes from identifying all the possible dilations of the channel as the primitive \blk object whose dynamics is to be studied. This formalism for channels, however, does not generalise straightforwardly to the case of dynamics of process matrices, since within the current notions of a Stinespring-like dilation of process matrices, not all process matrices have a unitary dilation \cite{araujo2017purification}. We hope that the formalism for channel dynamics that we propose will, however, provide inspiration to progress the study of process-matrix dynamics.

We see hence that in order to fully understand the dynamics of process matrices, one must go beyond the restrictive setup {{of predictable evolution of black boxes} that arises from higher-order processes} and consider other possible evolution mechanisms {which depend on the physical implementation of the process matrix} {(see Figure~\ref{fig:thefig})}. Our work is particularly relevant for the study of quantum networks, in particular to understand which resources are necessary to construct networks that feature an indefinite causal structure. A recent conjecture, vital to these issues, is that ``causally separable processes cannot evolve into causally non-separable ones via transformations that have only continuous, physical dilations'' \cite{castro2018dynamics}. Our results, in contrast, suggest that this might not be the case, and that if one is to consider a more comprehensive type of dynamics, then the opposite conclusion could be drawn. {Quite interestingly, though, this does not need to lead to a  worrying contradiction: as mentioned already in Section~\ref{sec:overview}, we may reconcile the whole picture by viewing our dynamics and those of Ref.~\cite{castro2018dynamics} as belonging to a hierarchy of possible, closed classes of dynamics acting on physical channels---in loose analogy to the well-known LO vs. LOCC distinction in entanglement theory (see Ref.~\cite{horodecki2009quantum}).}

{{Remarkably, our results now allow us to  provide a more refined answer to the {central} question: ``when can causal order continuously and reversibly evolve?''. This answer is} ``when we do not {focus on} being able to predict the evolution, or, when we have some knowledge of {the object's} physical implementation.''.}

\

{\it Authors Constributions:} Conceptualisation, J.H.S., A.B.S. and P.H.; Formal analysis, J.H.S., A.B.S. and P.H.; Investigation, J.H.S., A.B.S. and P.H.; Methodology, J.H.S., A.B.S. and P.H.; Writing---original draft, J.H.S., A.B.S. and P.H.; Writing---review and editing, J.H.S., A.B.S. and P.H. All authors have read and agreed to the published version of the manuscript.

\

{\it Funding:} A.B.S.~and P.H.~acknowledge support by the Foundation for Polish Science  (IRAP project, ICTQT, contract no.2018/MAB/5, co-financed by EU within Smart Growth Operational Programme). J.H.S.~was supported by the National Science Centre, Poland (Opus project, Categorical Foundations of the Non-Classicality of Nature, project no.~2021/41/B/ST2/03149).

\

{\it Acknowledgements:} We thank  Esteban Castro-Ruiz, Flaminia Giacomini, and
\v{C}aslav
Brukner for insightful discussions about their work which led to the development of section IV of this manuscript. JHS thanks David Schmid for interesting discussions in the early stages of this project. 
 All of the diagrams within this manuscript were prepared using TikZit.

\end{document}

%% file: preamble.tex
\usepackage{rotating}
\usepackage{floatpag}
\rotfloatpagestyle{empty}

\usepackage{hyperref}
\usepackage{dsfont}
\usepackage{amsmath,amsthm,amssymb}
\usepackage{xspace,enumerate,epsfig}
\usepackage{graphicx}
\graphicspath{{.}{./figures/}}
\usepackage{marvosym}

\usepackage{tikzfig}
\usepackage{stmaryrd}
\usepackage{docmute}
\usepackage{keycommand}

\input{defs.tex}

\input{tikzstyles.tex}

\let\olddagger\dagger
\renewcommand{\dagger}{\ensuremath{\olddagger}\xspace}

\theoremstyle{definition}

\newtheorem{example*}[theorem]{Example*}
\newtheorem{examples*}[theorem]{Examples*}

\newtheorem{remark*}[theorem]{Remark*}

\theoremstyle{plain}

\usepackage{color}
\def\bR{\begin{color}{red}}
\def\bB{\begin{color}{blue}}
\def\bM{\begin{color}{magenta}}
\def\bC{\begin{color}{cyan}}
\def\bW{\begin{color}{white}}
\def\bBl{\begin{color}{black}}
\def\bG{\begin{color}{green}}
\def\bY{\begin{color}{yellow}}
\def\e{\end{color}\xspace}
\newcommand{\bit}{\begin{itemize}}
\newcommand{\eit}{\end{itemize}\par\noindent}
\newcommand{\ben}{\begin{enumerate}}
\newcommand{\een}{\end{enumerate}\par\noindent}
\newcommand{\beq}{\begin{equation}}
\newcommand{\eeq}{\end{equation}\par\noindent}
\newcommand{\beqa}{\begin{eqnarray*}}
\newcommand{\eeqa}{\end{eqnarray*}\par\noindent}
\newcommand{\beqn}{\begin{eqnarray}}
\newcommand{\eeqn}{\end{eqnarray}\par\noindent}

\def\jR{\begin{color}{DarkRed}}
\def\jB{\begin{color}{MidnightBlue}}
\def\jM{\begin{color}{magenta}}
\def\jC{\begin{color}{cyan}}
\def\jW{\begin{color}{white}}
\def\jBl{\begin{color}{black}}
\def\jG{\begin{color}{green}}
\def\jY{\begin{color}{yellow}}

\def\cR{\begin{color}{Crimson}}
\def\cB{\begin{color}{cyan}}
\def\cM{\begin{color}{magenta}}
\def\cC{\begin{color}{cyan}}
\def\cW{\begin{color}{white}}
\def\cBl{\begin{color}{black}}
\def\cG{\begin{color}{green}}
\def\cY{\begin{color}{yellow}}


%% file: defs.tex
\newkeycommand{\pointmap}[style=point][1]{\,\tikz{\node[style=\commandkey{style}] (x) at (0,-0.3) {$#1$};\node [style=none] (3) at (0, 1.0) {};\node [style=none] (3) at (0, -1.0) {};\draw (x)--(0,0.7);}\,}

\newkeycommand{\typedkpoint}[style=kpoint,edgestyle=][2]{%
\,\begin{tikzpicture}
  \begin{pgfonlayer}{nodelayer}
    \node [style=none] (0) at (0, 1) {};
    \node [style=\commandkey{style}] (1) at (0, -0.75) {$#2$};
    \node [style=right label] (2) at (0.25, 0.5) {$#1$};
  \end{pgfonlayer}
  \begin{pgfonlayer}{edgelayer}
    \draw [\commandkey{edgestyle}] (1) to (0.center);
  \end{pgfonlayer}
\end{tikzpicture}\,}

\newkeycommand{\typedkpointdag}[style=kpoint adjoint,edgestyle=][2]{%
\,\begin{tikzpicture}
  \begin{pgfonlayer}{nodelayer}
    \node [style=none] (0) at (0, -1) {};
    \node [style=\commandkey{style}] (1) at (0, 0.75) {$#2$};
    \node [style=right label] (2) at (0.25, -0.5) {$#1$};
  \end{pgfonlayer}
  \begin{pgfonlayer}{edgelayer}
    \draw [\commandkey{edgestyle}] (0.center) to (1);
  \end{pgfonlayer}
\end{tikzpicture}\,}

\newcommand{\bistateadj}[1]{\,\begin{tikzpicture}[yshift=-3mm]
  \begin{pgfonlayer}{nodelayer}
    \node [style=kpointadj, minimum width=1 cm, inner sep=2pt] (0) at (0, 0.5) {$\psi$};
    \node [style=none] (1) at (-0.75, 0.25) {};
    \node [style=none] (2) at (0.75, 0.25) {};
    \node [style=none] (3) at (-0.75, -0.5) {};
    \node [style=none] (4) at (0.75, -0.5) {};
  \end{pgfonlayer}
  \begin{pgfonlayer}{edgelayer}
    \draw (1.center) to (3.center);
    \draw (2.center) to (4.center);
  \end{pgfonlayer}
\end{tikzpicture}\,}

\newkeycommand{\pointketbra}[style1=copoint,style2=point][2]{\,%
\begin{tikzpicture}
  \begin{pgfonlayer}{nodelayer}
    \node [style=none] (0) at (0, -1.5) {};
    \node [style=\commandkey{style1}] (1) at (0, -0.75) {$#1$};
    \node [style=\commandkey{style2}] (2) at (0, 0.75) {$#2$};
    \node [style=none] (3) at (0, 1.5) {};
  \end{pgfonlayer}
  \begin{pgfonlayer}{edgelayer}
    \draw (0.center) to (1);
    \draw (2) to (3.center);
  \end{pgfonlayer}
\end{tikzpicture}\,}

\newkeycommand{\twocopointketbra}[style1=copoint,style2=copoint,style3=point][3]{\,%
\begin{tikzpicture}
  \begin{pgfonlayer}{nodelayer}
    \node [style=none] (0) at (-0.75, -1.5) {};
    \node [style=none] (1) at (0.75, -1.5) {};
    \node [style=\commandkey{style1}] (2) at (-0.75, -0.75) {$#1$};
    \node [style=\commandkey{style2}] (3) at (0.75, -0.75) {$#2$};
    \node [style=\commandkey{style3}] (4) at (0, 0.75) {$#3$};
    \node [style=none] (5) at (0, 1.5) {};
  \end{pgfonlayer}
  \begin{pgfonlayer}{edgelayer}
    \draw (0.center) to (2);
    \draw (1.center) to (3);
    \draw (4) to (5.center);
  \end{pgfonlayer}
\end{tikzpicture}\,}

\newkeycommand{\twopointketbra}[style1=copoint,style2=point,style3=point][3]{\,%
\begin{tikzpicture}
  \begin{pgfonlayer}{nodelayer}
    \node [style=none] (0) at (0, -1.5) {};
    \node [style=none] (1) at (-0.75, 1.5) {};
    \node [style=\commandkey{style1}] (2) at (0, -0.75) {$#1$};
    \node [style=\commandkey{style2}] (3) at (-0.75, 0.75) {$#2$};
    \node [style=\commandkey{style3}] (4) at (0.75, 0.75) {$#3$};
    \node [style=none] (5) at (0.75, 1.5) {};
  \end{pgfonlayer}
  \begin{pgfonlayer}{edgelayer}
    \draw (0.center) to (2);
    \draw (1.center) to (3);
    \draw (4) to (5.center);
  \end{pgfonlayer}
\end{tikzpicture}\,}

\newkeycommand{\pointbraket}[style1=point,style2=copoint][2]{\,%
\begin{tikzpicture}
  \begin{pgfonlayer}{nodelayer}
    \node [style=\commandkey{style1}] (0) at (0, -0.5) {$#1$};
    \node [style=\commandkey{style2}] (1) at (0, 0.5) {$#2$};
  \end{pgfonlayer}
  \begin{pgfonlayer}{edgelayer}
    \draw (0) to (1);
  \end{pgfonlayer}
\end{tikzpicture}\,}

\newkeycommand{\kpointbraket}[style1=kpoint,style2=kpoint adjoint][2]{\,%
\begin{tikzpicture}
  \begin{pgfonlayer}{nodelayer}
    \node [style=\commandkey{style1}] (0) at (0, -0.75) {$#1$};
    \node [style=\commandkey{style2}] (1) at (0, 0.75) {$#2$};
  \end{pgfonlayer}
  \begin{pgfonlayer}{edgelayer}
    \draw (0) to (1);
  \end{pgfonlayer}
\end{tikzpicture}\,}

\newkeycommand{\kpointmap}[style1=kpoint,style2=map][2]{\,%
\begin{tikzpicture}
  \begin{pgfonlayer}{nodelayer}
    \node [style=\commandkey{style1}] (0) at (0, -1.1) {$#1$};
    \node [style=\commandkey{style2}] (1) at (0, 0.2) {$#2$};
    \node [style=none] (2) at (0, 1.5) {};
  \end{pgfonlayer}
  \begin{pgfonlayer}{edgelayer}
    \draw (0) to (1);
    \draw (1) to (2.center);
  \end{pgfonlayer}
\end{tikzpicture}%
\,}

\newkeycommand{\sandwichtwo}[style1=point,style2=map,style3=map,style4=copoint][4]{\,%
\begin{tikzpicture}
  \begin{pgfonlayer}{nodelayer}
    \node [style=\commandkey{style2}] (0) at (0, -0.75) {$#2$};
    \node [style=\commandkey{style3}] (1) at (0, 0.75) {$#3$};
    \node [style=\commandkey{style1}] (2) at (0, -2) {$#1$};
    \node [style=\commandkey{style4}] (3) at (0, 2) {$#4$};
  \end{pgfonlayer}
  \begin{pgfonlayer}{edgelayer}
    \draw (2) to (0);
    \draw (0) to (1);
    \draw (1) to (3);
  \end{pgfonlayer}
\end{tikzpicture}\,}

\newkeycommand{\longbraket}[style1=point,style2=copoint][2]{\,%
\begin{tikzpicture}
  \begin{pgfonlayer}{nodelayer}
    \node [style=\commandkey{style1}] (0) at (0, -2) {$#1$};
    \node [style=\commandkey{style2}] (1) at (0, 2) {$#2$};
  \end{pgfonlayer}
  \begin{pgfonlayer}{edgelayer}
    \draw (0) to (1);
  \end{pgfonlayer}
\end{tikzpicture}\,}

\newkeycommand{\onb}[style=point]{\ensuremath{\left\{\tikz{\node[style=\commandkey{style}] (x) at (0,-0.3) {$j$};\draw (x)--(0,0.7);}\right\}}\xspace}

\newkeycommand{\copointmap}[style=copoint][1]{\,\tikz{\node[style=\commandkey{style}] (x) at (0,0.3) {$#1$};\draw (x)--(0,-0.7);}\,}

\newcommand{\bra}[1]{\ensuremath{\left\langle #1 \right|}}
\newcommand{\ket}[1]{\ensuremath{\left|  #1 \right\rangle}}

\tikzstyle{cdiag}=[matrix of math nodes, row sep=3em, column sep=3em, text height=1.5ex, text depth=0.25ex,inner sep=0.5em]
\tikzstyle{arrow above}=[transform canvas={yshift=0.5ex}]
\tikzstyle{arrow below}=[transform canvas={yshift=-0.5ex}]



%% file: tikzstyles.tex
\usetikzlibrary{shapes.multipart}

\tikzstyle{bwSpider}=[
       rectangle split,
       rectangle split parts=2,
       rectangle split part fill={black,white},
 minimum size=3.6 mm, inner sep=-2mm, draw=black,scale=0.5,rounded corners=0.8 mm
       ]
 \tikzstyle{wbSpider}=[
       rectangle split,
       rectangle split parts=2,
       rectangle split part fill={white,black},
 minimum size=3.6 mm, inner sep=-2mm, draw=black,scale=0.5,rounded corners=0.8 mm
       ]
\tikzstyle{qWire}=[line width = 1pt, color=black]
\tikzstyle{cWire}=[color=quantumgray,thin]
\tikzstyle{env}=[copoint,regular polygon rotate=0,minimum width=0.2cm, fill=black]

\tikzstyle{probs}=[shape=semicircle,fill=white,draw=black,shape border rotate=180,minimum width=1.2cm]

\tikzstyle{every picture}=[baseline=-0.25em,scale=0.5]
\tikzstyle{dotpic}=[] 
\tikzstyle{diredges}=[every to/.style={diredge}]
\tikzstyle{math matrix}=[matrix of math nodes,left delimiter=(,right delimiter=),inner sep=2pt,column sep=1em,row sep=0.5em,nodes={inner sep=0pt},text height=1.5ex, text depth=0.25ex]

\tikzstyle{inline text}=[text height=1.5ex, text depth=0.25ex,yshift=0.5mm]
\tikzstyle{label}=[font=\scriptsize,text height=1.5ex, text depth=0.25ex,yshift=0.5mm]
\tikzstyle{left label}=[label,anchor=east,xshift=1.5mm]
\tikzstyle{right label}=[label,anchor=west,xshift=-1mm]

\tikzstyle{braceedge}=[decorate,decoration={brace,amplitude=2mm,raise=-1mm}]
\tikzstyle{small braceedge}=[decorate,decoration={brace,amplitude=1mm,raise=-1mm}]

\tikzstyle{doubled}=[line width=1.6pt] 
\tikzstyle{boldedge}=[doubled,shorten <=-0.17mm,shorten >=-0.17mm]
\tikzstyle{boldedgegray}=[doubled,gray,shorten <=-0.17mm,shorten >=-0.17mm]
\tikzstyle{singleedgegray}=[gray]

\tikzstyle{semidoubled}=[line width=1.4pt] 
\tikzstyle{semiboldedgegray}=[semidoubled,gray,shorten <=-0.17mm,shorten >=-0.17mm]

\tikzstyle{boxedge}=[semiboldedgegray]

\tikzstyle{boldedgedashed}=[very thick,dashed,shorten <=-0.17mm,shorten >=-0.17mm]
\tikzstyle{vboldedgedashed}=[doubled,dashed,shorten <=-0.17mm,shorten >=-0.17mm]
\tikzstyle{left hook arrow}=[left hook-latex]
\tikzstyle{right hook arrow}=[right hook-latex]
\tikzstyle{sembracket}=[line width=0.5pt,shorten <=-0.07mm,shorten >=-0.07mm]

\tikzstyle{causal edge}=[->,thick,gray]
\tikzstyle{causal nondir}=[thick,gray]
\tikzstyle{timeline}=[thick,gray, dashed]

\tikzstyle{cedge}=[<->,thick,gray!70!white]

\tikzstyle{empty diagram}=[draw=gray!40!white,dashed,shape=rectangle,minimum width=1cm,minimum height=1cm]
\tikzstyle{empty diagram small}=[draw=gray!50!white,dashed,shape=rectangle,minimum width=0.6cm,minimum height=0.5cm]

\tikzstyle{dot}=[inner sep=0mm,minimum width=2mm,minimum height=2mm,draw,shape=circle]
\tikzstyle{leak}=[white dot, shape=regular polygon, minimum size=3.3 mm, regular polygon sides=3, outer sep=-0.2mm, regular polygon rotate=270]
\tikzstyle{proj}=[draw,fill=white,chamfered rectangle,chamfered rectangle angle=30, minimum width=2mm, minimum height= 1mm,scale=0.5, outer sep=-0.2mm]
\tikzstyle{wide proj}=[draw,fill=white,chamfered rectangle,chamfered rectangle angle=30, minimum width=15mm,minimum height=1mm,scale=0.5, outer sep=-0.2mm]
\tikzstyle{very wide proj}=[draw,fill=white,chamfered rectangle,chamfered rectangle angle=30, minimum width=25mm,minimum height=1mm,scale=0.5, outer sep=-0.2mm]
\tikzstyle{very very wide proj}=[draw,fill=white,chamfered rectangle,chamfered rectangle angle=30, minimum width=35mm,minimum height=1mm,scale=0.5, outer sep=-0.2mm]
\tikzstyle{preleak}=[proj]

\tikzstyle{split proj out}=[regular polygon,regular polygon sides=3,draw,scale=0.75,inner sep=-0.5pt,minimum width=3.3mm,fill=white,regular polygon rotate=180]
\tikzstyle{split proj in}=[regular polygon,regular polygon sides=3,draw,scale=0.75,inner sep=-0.5pt,minimum width=3.3mm,fill=white]
\tikzstyle{Vleak}=[white dot, shape=regular polygon, minimum size=3.3 mm, regular polygon sides=3, outer sep=-0.2mm, regular polygon rotate=90]
\tikzstyle{dleak}=[white dot, line width=1.6pt, shape=regular polygon, minimum size=3.3 mm, regular polygon sides=3, outer sep=-0.2mm, regular polygon rotate=270]

\tikzstyle{Wsquare}=[white dot, shape=regular polygon, rounded corners=0.8 mm, minimum size=3.3 mm, regular polygon sides=3, outer sep=-0.2mm]
\tikzstyle{Wsquareadj}=[white dot, shape=regular polygon, rounded corners=0.8 mm, minimum size=3.3 mm, regular polygon sides=3, outer sep=-0.2mm, regular polygon rotate=180]
\tikzstyle{ddot}=[inner sep=0mm, doubled, minimum width=2.5mm,minimum height=2.5mm,draw,shape=circle]

\tikzstyle{black dot}=[dot,fill=black]
\tikzstyle{white dot}=[dot,fill=white,,text depth=-0.2mm]
\tikzstyle{white Wsquare}=[Wsquare,fill=gray,,text depth=-0.2mm]
\tikzstyle{white Wsquareadj}=[Wsquareadj,fill=white,,text depth=-0.2mm]
\tikzstyle{green dot}=[white dot] 
\tikzstyle{gray dot}=[dot,fill=gray!40!white,,text depth=-0.2mm]
\tikzstyle{red dot}=[gray dot] 

\tikzstyle{black ddot}=[ddot,fill=black]
\tikzstyle{white ddot}=[ddot,fill=white]
\tikzstyle{gray ddot}=[ddot,fill=gray!40!white]

\tikzstyle{gray edge}=[gray!60!white]

\tikzstyle{small dot}=[inner sep=0.5mm,minimum width=0pt,minimum height=0pt,draw,shape=circle]

\tikzstyle{small black dot}=[small dot,fill=black]
\tikzstyle{small white dot}=[small dot,fill=white]
\tikzstyle{small gray dot}=[small dot,fill=gray!40!white]

\tikzstyle{causal dot}=[inner sep=0.4mm,minimum width=0pt,minimum height=0pt,draw=white,shape=circle,fill=gray!40!white]

\tikzstyle{phase dimensions}=[minimum size=5mm,font=\footnotesize,rectangle,rounded corners=2.5mm,inner sep=0.2mm,outer sep=-2mm]
\tikzstyle{dphase dimensions}=[minimum size=5mm,font=\footnotesize,rectangle,rounded corners=2.5mm,inner sep=0.2mm,outer sep=-2mm]

\tikzstyle{white phase dot}=[dot,fill=white,phase dimensions]
\tikzstyle{white phase ddot}=[ddot,fill=white,dphase dimensions]

\tikzstyle{white rect ddot}=[draw=black,fill=white,doubled,minimum size=5mm,font=\footnotesize,rectangle,rounded corners=2.5mm,inner sep=0.2mm]
\tikzstyle{gray rect ddot}=[draw=black,fill=gray!40!white,doubled,minimum size=6mm,font=\footnotesize,rectangle,rounded corners=3mm]

\tikzstyle{gray phase dot}=[dot,fill=gray!40!white,phase dimensions]
\tikzstyle{gray phase ddot}=[ddot,fill=gray!40!white,dphase dimensions]
\tikzstyle{grey phase dot}=[gray phase dot]
\tikzstyle{grey phase ddot}=[gray phase ddot]

\tikzstyle{small phase dimensions}=[minimum size=4mm,font=\tiny,rectangle,rounded corners=2mm,inner sep=0.2mm,outer sep=-2mm]
\tikzstyle{small dphase dimensions}=[minimum size=4mm,font=\tiny,rectangle,rounded corners=2mm,inner sep=0.2mm,outer sep=-2mm]

\tikzstyle{small gray phase dot}=[dot,fill=gray!40!white,small phase dimensions]
\tikzstyle{small gray phase ddot}=[ddot,fill=gray!40!white,small dphase dimensions]

\tikzstyle{small map}=[draw,shape=rectangle,minimum height=4mm,minimum width=4mm,fill=white]

\tikzstyle{cnot}=[fill=white,shape=circle,inner sep=-1.4pt]

\tikzstyle{asym hadamard}=[fill=white,draw,shape=NEbox,inner sep=0.6mm,font=\footnotesize,minimum height=4mm]
\tikzstyle{asym hadamard conj}=[fill=white,draw,shape=NWbox,inner sep=0.6mm,font=\footnotesize,minimum height=4mm]
\tikzstyle{asym hadamard dag}=[fill=white,draw,shape=SEbox,inner sep=0.6mm,font=\footnotesize,minimum height=4mm]

\tikzstyle{hadamard}=[fill=white,draw,inner sep=0.6mm,font=\footnotesize,minimum height=4mm,minimum width=4mm]
\tikzstyle{small hadamard}=[fill=white,draw,inner sep=0.6mm,minimum height=1.5mm,minimum width=1.5mm]
\tikzstyle{small hadamard rotate}=[small hadamard,rotate=45]
\tikzstyle{dhadamard}=[hadamard,doubled]
\tikzstyle{small dhadamard}=[small hadamard,doubled]
\tikzstyle{small dhadamard rotate}=[small hadamard rotate,doubled]
\tikzstyle{antipode}=[white dot,inner sep=0.3mm,font=\footnotesize]

\tikzstyle{scalar}=[diamond,draw,inner sep=0.5pt,font=\small]
\tikzstyle{dscalar}=[diamond,doubled, draw,inner sep=0.5pt,font=\small]

\tikzstyle{small box}=[rectangle,inline text,fill=white,draw,minimum height=5mm,yshift=-0.5mm,minimum width=5mm,font=\small]
\tikzstyle{small gray box}=[small box,fill=gray!30]
\tikzstyle{medium box}=[rectangle,inline text,fill=white,draw,minimum height=5mm,yshift=-0.5mm,minimum width=10mm,font=\small]
\tikzstyle{square box}=[small box] 
\tikzstyle{medium gray box}=[small box,fill=gray!30]
\tikzstyle{semilarge box}=[rectangle,inline text,fill=white,draw,minimum height=5mm,yshift=-0.5mm,minimum width=12.5mm,font=\small]
\tikzstyle{large box}=[rectangle,inline text,fill=white,draw,minimum height=5mm,yshift=-0.5mm,minimum width=15mm,font=\small]
\tikzstyle{large gray box}=[small box,fill=gray!30]

\tikzstyle{Bayes box}=[rectangle,fill=black,draw, minimum height=3mm, minimum width=3mm]

\tikzstyle{gray square point}=[small box,fill=gray!50]

\tikzstyle{dphase box white}=[dhadamard]
\tikzstyle{dphase box gray}=[dhadamard,fill=gray!50!white]
\tikzstyle{phase box white}=[hadamard]
\tikzstyle{phase box gray}=[hadamard,fill=gray!50!white]

\tikzstyle{point}=[regular polygon,regular polygon sides=3,draw,scale=0.75,inner sep=-0.5pt,minimum width=9mm,fill=white,regular polygon rotate=180]
\tikzstyle{point nosep}=[regular polygon,regular polygon sides=3,draw,scale=0.75,inner sep=-2pt,minimum width=9mm,fill=white,regular polygon rotate=180]
\tikzstyle{copoint}=[regular polygon,regular polygon sides=3,draw,scale=0.75,inner sep=-0.5pt,minimum width=9mm,fill=white]
\tikzstyle{dpoint}=[point,doubled]
\tikzstyle{dcopoint}=[copoint,doubled]

\tikzstyle{pointgrow}=[shape=cornerpoint,kpoint common,scale=0.75,inner sep=3pt]
\tikzstyle{pointgrow dag}=[shape=cornercopoint,kpoint common,scale=0.75,inner sep=3pt]

\tikzstyle{wide copoint}=[fill=white,draw,shape=isosceles triangle,shape border rotate=90,isosceles triangle stretches=true,inner sep=0pt,minimum width=1.5cm,minimum height=6.12mm]
\tikzstyle{wide point}=[fill=white,draw,shape=isosceles triangle,shape border rotate=-90,isosceles triangle stretches=true,inner sep=0pt,minimum width=1.5cm,minimum height=6.12mm,yshift=-0.0mm]
\tikzstyle{wide point plus}=[fill=white,draw,shape=isosceles triangle,shape border rotate=-90,isosceles triangle stretches=true,inner sep=0pt,minimum width=1.74cm,minimum height=7mm,yshift=-0.0mm]

\tikzstyle{wide dpoint}=[fill=white,doubled,draw,shape=isosceles triangle,shape border rotate=-90,isosceles triangle stretches=true,inner sep=0pt,minimum width=1.5cm,minimum height=6.12mm,yshift=-0.0mm]

\tikzstyle{tinypoint}=[regular polygon,regular polygon sides=3,draw,scale=0.55,inner sep=-0.15pt,minimum width=6mm,fill=white,regular polygon rotate=180]

\tikzstyle{white point}=[point]
\tikzstyle{white dpoint}=[dpoint]
\tikzstyle{green point}=[white point] 
\tikzstyle{white copoint}=[copoint]
\tikzstyle{gray point}=[point,fill=gray!40!white]
\tikzstyle{gray dpoint}=[gray point,doubled]
\tikzstyle{red point}=[gray point] 
\tikzstyle{gray copoint}=[copoint,fill=gray!40!white]
\tikzstyle{gray dcopoint}=[gray copoint,doubled]

\tikzstyle{white point guide}=[regular polygon,regular polygon sides=3,font=\scriptsize,draw,scale=0.65,inner sep=-0.5pt,minimum width=9mm,fill=white,regular polygon rotate=180]

\tikzstyle{black point}=[point,fill=black,font=\color{white}]
\tikzstyle{black copoint}=[copoint,fill=black,font=\color{white}]

\tikzstyle{tiny gray point}=[tinypoint,fill=gray!40!white]

\tikzstyle{diredge}=[->]
\tikzstyle{ddiredge}=[<->]
\tikzstyle{rdiredge}=[<-]
\tikzstyle{thickdiredge}=[->, very thick]
\tikzstyle{pointer edge}=[->,very thick,gray]
\tikzstyle{pointer edge part}=[very thick,gray]
\tikzstyle{dashed edge}=[dashed]
\tikzstyle{thick dashed edge}=[very thick,dashed]
\tikzstyle{thick gray dashed edge}=[thick dashed edge,gray!40]
\tikzstyle{thick map edge}=[very thick,|->]

\makeatletter
\newcommand{\boxshape}[3]{%
\pgfdeclareshape{#1}{
\inheritsavedanchors[from=rectangle] 
\inheritanchorborder[from=rectangle]
\inheritanchor[from=rectangle]{center}
\inheritanchor[from=rectangle]{north}
\inheritanchor[from=rectangle]{south}
\inheritanchor[from=rectangle]{west}
\inheritanchor[from=rectangle]{east}
\backgroundpath{
\southwest \pgf@xa=\pgf@x \pgf@ya=\pgf@y
\northeast \pgf@xb=\pgf@x \pgf@yb=\pgf@y

\@tempdima=#2
\@tempdimb=#3

\pgfpathmoveto{\pgfpoint{\pgf@xa - 5pt + \@tempdima}{\pgf@ya}}
\pgfpathlineto{\pgfpoint{\pgf@xa - 5pt - \@tempdima}{\pgf@yb}}
\pgfpathlineto{\pgfpoint{\pgf@xb + 5pt + \@tempdimb}{\pgf@yb}}
\pgfpathlineto{\pgfpoint{\pgf@xb + 5pt - \@tempdimb}{\pgf@ya}}
\pgfpathlineto{\pgfpoint{\pgf@xa - 5pt + \@tempdima}{\pgf@ya}}
\pgfpathclose
}
}}

\boxshape{NEbox}{0pt}{3pt}
\boxshape{SEbox}{0pt}{-3pt}
\boxshape{NWbox}{5pt}{0pt}
\boxshape{SWbox}{-5pt}{0pt}
\boxshape{EBox}{-3pt}{3pt}
\boxshape{WBox}{3pt}{-3pt}
\makeatother

\tikzstyle{cloud}=[shape=cloud,draw,minimum width=1.5cm,minimum height=1.5cm]

\tikzstyle{map}=[draw,shape=NEbox,inner sep=1pt,minimum height=4mm,fill=white]
\tikzstyle{dashedmap}=[draw,dashed,shape=NEbox,inner sep=2pt,minimum height=6mm,fill=white]
\tikzstyle{mapdag}=[draw,shape=SEbox,inner sep=1pt,minimum height=4mm,fill=white]
\tikzstyle{mapadj}=[draw,shape=SEbox,inner sep=2pt,minimum height=6mm,fill=white]
\tikzstyle{maptrans}=[draw,shape=SWbox,inner sep=2pt,minimum height=6mm,fill=white]
\tikzstyle{mapconj}=[draw,shape=NWbox,inner sep=2pt,minimum height=6mm,fill=white]

\tikzstyle{medium map}=[draw,shape=NEbox,inner sep=2pt,minimum height=6mm,fill=white,minimum width=7mm]
\tikzstyle{medium map dag}=[draw,shape=SEbox,inner sep=2pt,minimum height=6mm,fill=white,minimum width=7mm]
\tikzstyle{medium map adj}=[draw,shape=SEbox,inner sep=2pt,minimum height=6mm,fill=white,minimum width=7mm]
\tikzstyle{medium map trans}=[draw,shape=SWbox,inner sep=2pt,minimum height=6mm,fill=white,minimum width=7mm]
\tikzstyle{medium map conj}=[draw,shape=NWbox,inner sep=2pt,minimum height=6mm,fill=white,minimum width=7mm]
\tikzstyle{semilarge map}=[draw,shape=NEbox,inner sep=2pt,minimum height=6mm,fill=white,minimum width=9.5mm]
\tikzstyle{semilarge map trans}=[draw,shape=SWbox,inner sep=2pt,minimum height=6mm,fill=white,minimum width=9.5mm]
\tikzstyle{semilarge map adj}=[draw,shape=SEbox,inner sep=2pt,minimum height=6mm,fill=white,minimum width=9.5mm]
\tikzstyle{semilarge map dag}=[draw,shape=SEbox,inner sep=2pt,minimum height=6mm,fill=white,minimum width=9.5mm]
\tikzstyle{semilarge map conj}=[draw,shape=NWbox,inner sep=2pt,minimum height=6mm,fill=white,minimum width=9.5mm]
\tikzstyle{large map}=[draw,shape=NEbox,inner sep=2pt,minimum height=6mm,fill=white,minimum width=12mm]
\tikzstyle{large map conj}=[draw,shape=NWbox,inner sep=2pt,minimum height=6mm,fill=white,minimum width=12mm]
\tikzstyle{very large map}=[draw,shape=NEbox,inner sep=2pt,minimum height=6mm,fill=white,minimum width=17mm]

\tikzstyle{medium dmap}=[draw,doubled,shape=NEbox,inner sep=2pt,minimum height=6mm,fill=white,minimum width=7mm]
\tikzstyle{medium dmap dag}=[draw,doubled,shape=SEbox,inner sep=2pt,minimum height=6mm,fill=white,minimum width=7mm]
\tikzstyle{medium dmap adj}=[draw,doubled,shape=SEbox,inner sep=2pt,minimum height=6mm,fill=white,minimum width=7mm]
\tikzstyle{medium dmap trans}=[draw,doubled,shape=SWbox,inner sep=2pt,minimum height=6mm,fill=white,minimum width=7mm]
\tikzstyle{medium dmap conj}=[draw,doubled,shape=NWbox,inner sep=2pt,minimum height=6mm,fill=white,minimum width=7mm]
\tikzstyle{semilarge dmap}=[draw,doubled,shape=NEbox,inner sep=2pt,minimum height=6mm,fill=white,minimum width=9.5mm]
\tikzstyle{semilarge dmap trans}=[draw,doubled,shape=SWbox,inner sep=2pt,minimum height=6mm,fill=white,minimum width=9.5mm]
\tikzstyle{semilarge dmap adj}=[draw,doubled,shape=SEbox,inner sep=2pt,minimum height=6mm,fill=white,minimum width=9.5mm]
\tikzstyle{semilarge dmap dag}=[draw,doubled,shape=SEbox,inner sep=2pt,minimum height=6mm,fill=white,minimum width=9.5mm]
\tikzstyle{semilarge dmap conj}=[draw,doubled,shape=NWbox,inner sep=2pt,minimum height=6mm,fill=white,minimum width=9.5mm]
\tikzstyle{large dmap}=[draw,doubled,shape=NEbox,inner sep=2pt,minimum height=6mm,fill=white,minimum width=12mm]
\tikzstyle{large dmap conj}=[draw,doubled,shape=NWbox,inner sep=2pt,minimum height=6mm,fill=white,minimum width=12mm]
\tikzstyle{large dmap trans}=[draw,doubled,shape=SWbox,inner sep=2pt,minimum height=6mm,fill=white,minimum width=12mm]
\tikzstyle{large dmap adj}=[draw,doubled,shape=SEbox,inner sep=2pt,minimum height=6mm,fill=white,minimum width=12mm]
\tikzstyle{large dmap dag}=[draw,doubled,shape=SEbox,inner sep=2pt,minimum height=6mm,fill=white,minimum width=12mm]
\tikzstyle{very large dmap}=[draw,doubled,shape=NEbox,inner sep=2pt,minimum height=6mm,fill=white,minimum width=19.5mm]

\tikzstyle{muxbox}=[draw,shape=rectangle,minimum height=3mm,minimum width=3mm,fill=white]
\tikzstyle{dmuxbox}=[muxbox,doubled]

\tikzstyle{box}=[draw,shape=rectangle,inner sep=2pt,minimum height=6mm,minimum width=6mm,fill=white]
\tikzstyle{dbox}=[draw,doubled,shape=rectangle,inner sep=2pt,minimum height=6mm,minimum width=6mm,fill=white]
\tikzstyle{dmap}=[draw,doubled,shape=NEbox,inner sep=2pt,minimum height=6mm,fill=white]
\tikzstyle{dmapdag}=[draw,doubled,shape=SEbox,inner sep=2pt,minimum height=6mm,fill=white]
\tikzstyle{dmapadj}=[draw,doubled,shape=SEbox,inner sep=2pt,minimum height=6mm,fill=white]
\tikzstyle{dmaptrans}=[draw,doubled,shape=SWbox,inner sep=2pt,minimum height=6mm,fill=white]
\tikzstyle{dmapconj}=[draw,doubled,shape=NWbox,inner sep=2pt,minimum height=6mm,fill=white]

\tikzstyle{ddmap}=[draw,doubled,dashed,shape=NEbox,inner sep=2pt,minimum height=6mm,fill=white]
\tikzstyle{ddmapdag}=[draw,doubled,dashed,shape=SEbox,inner sep=2pt,minimum height=6mm,fill=white]
\tikzstyle{ddmapadj}=[draw,doubled,dashed,shape=SEbox,inner sep=2pt,minimum height=6mm,fill=white]
\tikzstyle{ddmaptrans}=[draw,doubled,dashed,shape=SWbox,inner sep=2pt,minimum height=6mm,fill=white]
\tikzstyle{ddmapconj}=[draw,doubled,dashed,shape=NWbox,inner sep=2pt,minimum height=6mm,fill=white]

\boxshape{sNEbox}{0pt}{3pt}
\boxshape{sSEbox}{0pt}{-3pt}
\boxshape{sNWbox}{3pt}{0pt}
\boxshape{sSWbox}{-3pt}{0pt}
\tikzstyle{smap}=[draw,shape=sNEbox,fill=white]
\tikzstyle{smapdag}=[draw,shape=sSEbox,fill=white]
\tikzstyle{smapadj}=[draw,shape=sSEbox,fill=white]
\tikzstyle{smaptrans}=[draw,shape=sSWbox,fill=white]
\tikzstyle{smapconj}=[draw,shape=sNWbox,fill=white]

\tikzstyle{dsmap}=[draw,dashed,shape=sNEbox,fill=white]
\tikzstyle{dsmapdag}=[draw,dashed,shape=sSEbox,fill=white]
\tikzstyle{dsmaptrans}=[draw,dashed,shape=sSWbox,fill=white]
\tikzstyle{dsmapconj}=[draw,dashed,shape=sNWbox,fill=white]

\boxshape{mNEbox}{0pt}{10pt}
\boxshape{mSEbox}{0pt}{-10pt}
\boxshape{mNWbox}{10pt}{0pt}
\boxshape{mSWbox}{-10pt}{0pt}
\tikzstyle{mmap}=[draw,shape=mNEbox]
\tikzstyle{mmapdag}=[draw,shape=mSEbox]
\tikzstyle{mmaptrans}=[draw,shape=mSWbox]
\tikzstyle{mmapconj}=[draw,shape=mNWbox]

\tikzstyle{mmapgray}=[draw,fill=gray!40!white,shape=mNEbox]
\tikzstyle{smapgray}=[draw,fill=gray!40!white,shape=sNEbox]

\makeatletter

\pgfdeclareshape{cornerpoint}{
\inheritsavedanchors[from=rectangle] 
\inheritanchorborder[from=rectangle]
\inheritanchor[from=rectangle]{center}
\inheritanchor[from=rectangle]{north}
\inheritanchor[from=rectangle]{south}
\inheritanchor[from=rectangle]{west}
\inheritanchor[from=rectangle]{east}
\backgroundpath{
\southwest \pgf@xa=\pgf@x \pgf@ya=\pgf@y
\northeast \pgf@xb=\pgf@x \pgf@yb=\pgf@y

\pgfmathsetmacro{\pgf@shorten@left}{\pgfkeysvalueof{/tikz/shorten left}}
\pgfmathsetmacro{\pgf@shorten@right}{\pgfkeysvalueof{/tikz/shorten right}}

\pgfpathmoveto{\pgfpoint{0.5 * (\pgf@xa + \pgf@xb)}{\pgf@ya - 5pt}}
\pgfpathlineto{\pgfpoint{\pgf@xa - 8pt + \pgf@shorten@left}{\pgf@yb - 1.5 * \pgf@shorten@left}}
\pgfpathlineto{\pgfpoint{\pgf@xa - 8pt + \pgf@shorten@left}{\pgf@yb}}
\pgfpathlineto{\pgfpoint{\pgf@xb + 8pt - \pgf@shorten@right}{\pgf@yb}}
\pgfpathlineto{\pgfpoint{\pgf@xb + 8pt - \pgf@shorten@right}{\pgf@yb - 1.5 * \pgf@shorten@right}}
\pgfpathclose
}
}

\pgfdeclareshape{cornercopoint}{
\inheritsavedanchors[from=rectangle] 
\inheritanchorborder[from=rectangle]
\inheritanchor[from=rectangle]{center}
\inheritanchor[from=rectangle]{north}
\inheritanchor[from=rectangle]{south}
\inheritanchor[from=rectangle]{west}
\inheritanchor[from=rectangle]{east}
\backgroundpath{
\southwest \pgf@xa=\pgf@x \pgf@ya=\pgf@y
\northeast \pgf@xb=\pgf@x \pgf@yb=\pgf@y

\pgfmathsetmacro{\pgf@shorten@left}{\pgfkeysvalueof{/tikz/shorten left}}
\pgfmathsetmacro{\pgf@shorten@right}{\pgfkeysvalueof{/tikz/shorten right}}

\pgfpathmoveto{\pgfpoint{0.5 * (\pgf@xa + \pgf@xb)}{\pgf@yb + 5pt}}
\pgfpathlineto{\pgfpoint{\pgf@xa - 8pt + \pgf@shorten@left}{\pgf@ya + 1.5 * \pgf@shorten@left}}
\pgfpathlineto{\pgfpoint{\pgf@xa - 8pt + \pgf@shorten@left}{\pgf@ya}}
\pgfpathlineto{\pgfpoint{\pgf@xb + 8pt - \pgf@shorten@right}{\pgf@ya}}
\pgfpathlineto{\pgfpoint{\pgf@xb + 8pt - \pgf@shorten@right}{\pgf@ya + 1.5 * \pgf@shorten@right}}
\pgfpathclose
}
}

\makeatother

\pgfkeyssetvalue{/tikz/shorten left}{0pt}
\pgfkeyssetvalue{/tikz/shorten right}{0pt}

\tikzstyle{kpoint common}=[draw,fill=white,inner sep=1pt,minimum height=4mm]
\tikzstyle{kpoint sc}=[shape=cornerpoint,kpoint common]
\tikzstyle{kpoint adjoint sc}=[shape=cornercopoint,kpoint common]
\tikzstyle{kpoint}=[shape=cornerpoint,shorten left=5pt,kpoint common]
\tikzstyle{kpoint adjoint}=[shape=cornercopoint,shorten left=5pt,kpoint common]
\tikzstyle{kpoint conjugate}=[shape=cornerpoint,shorten right=5pt,kpoint common]
\tikzstyle{kpoint transpose}=[shape=cornercopoint,shorten right=5pt,kpoint common]
\tikzstyle{kpoint symm}=[shape=cornerpoint,shorten left=5pt,shorten right=5pt,kpoint common]

\tikzstyle{wide kpoint sc}=[shape=cornerpoint,kpoint common, minimum width=1 cm]
\tikzstyle{wide kpointdag sc}=[shape=cornercopoint,kpoint common, minimum width=1 cm]

\tikzstyle{black kpoint}=[shape=cornerpoint,shorten left=5pt,kpoint common,fill=black,font=\color{white}]

\tikzstyle{black kpoint sm}=[shape=cornerpoint,shorten left=5pt,kpoint common,fill=black,font=\color{white},scale=0.75]

\tikzstyle{black kpoint adjoint}=[shape=cornercopoint,shorten left=5pt,kpoint common,fill=black,font=\color{white}]
\tikzstyle{black kpointadj}=[shape=cornercopoint,shorten left=5pt,kpoint common,fill=black,font=\color{white}]

\tikzstyle{black kpointadj sm}=[shape=cornercopoint,shorten left=5pt,kpoint common,fill=black,font=\color{white},scale=0.75]

\tikzstyle{black dkpoint}=[shape=cornerpoint,shorten left=5pt,kpoint common,fill=black, doubled,font=\color{white}]
\tikzstyle{black dkpoint adjoint}=[shape=cornercopoint,shorten left=5pt,kpoint common,fill=black, doubled,font=\color{white}]
\tikzstyle{black dkpointadj}=[shape=cornercopoint,shorten left=5pt,kpoint common,fill=black, doubled,font=\color{white}]

\tikzstyle{black dkpoint sm}=[shape=cornerpoint,shorten left=5pt,kpoint common,fill=black, doubled,font=\color{white},scale=0.75]
\tikzstyle{black dkpointadj sm}=[shape=cornercopoint,shorten left=5pt,kpoint common,fill=black, doubled,font=\color{white},scale=0.75]

\tikzstyle{kpointdag}=[kpoint adjoint]
\tikzstyle{kpointadj}=[kpoint adjoint]
\tikzstyle{kpointconj}=[kpoint conjugate]
\tikzstyle{kpointtrans}=[kpoint transpose]

\tikzstyle{big kpoint}=[kpoint, minimum width=1.2 cm, minimum height=8mm, inner sep=4pt, text depth=3mm]

\tikzstyle{wide kpoint}=[kpoint, minimum width=1 cm, inner sep=2pt]
\tikzstyle{wide kpointdag}=[kpointdag, minimum width=1 cm, inner sep=2pt]
\tikzstyle{wide kpointconj}=[kpointconj, minimum width=1 cm, inner sep=2pt]
\tikzstyle{wide kpointtrans}=[kpointtrans, minimum width=1 cm, inner sep=2pt]

\tikzstyle{wider kpoint}=[kpoint, minimum width=1.25 cm, inner sep=2pt]
\tikzstyle{wider kpointdag}=[kpointdag, minimum width=1.25 cm, inner sep=2pt]
\tikzstyle{wider kpointconj}=[kpointconj, minimum width=1.25 cm, inner sep=2pt]
\tikzstyle{wider kpointtrans}=[kpointtrans, minimum width=1.25 cm, inner sep=2pt]

\tikzstyle{gray kpoint}=[kpoint,fill=gray!50!white]
\tikzstyle{gray kpointdag}=[kpointdag,fill=gray!50!white]
\tikzstyle{gray kpointadj}=[kpointadj,fill=gray!50!white]
\tikzstyle{gray kpointconj}=[kpointconj,fill=gray!50!white]
\tikzstyle{gray kpointtrans}=[kpointtrans,fill=gray!50!white]

\tikzstyle{gray dkpoint}=[kpoint,fill=gray!50!white,doubled]
\tikzstyle{gray dkpointdag}=[kpointdag,fill=gray!50!white,doubled]
\tikzstyle{gray dkpointadj}=[kpointadj,fill=gray!50!white,doubled]
\tikzstyle{gray dkpointconj}=[kpointconj,fill=gray!50!white,doubled]
\tikzstyle{gray dkpointtrans}=[kpointtrans,fill=gray!50!white,doubled]

\tikzstyle{white label}=[draw,fill=white,rectangle,inner sep=0.7 mm]
\tikzstyle{gray label}=[draw,fill=gray!50!white,rectangle,inner sep=0.7 mm]
\tikzstyle{black label}=[draw,fill=black,rectangle,inner sep=0.7 mm]

\tikzstyle{dkpoint}=[kpoint,doubled]
\tikzstyle{wide dkpoint}=[wide kpoint,doubled]
\tikzstyle{dkpointdag}=[kpoint adjoint,doubled]
\tikzstyle{wide dkpointdag}=[wide kpointdag,doubled]
\tikzstyle{dkcopoint}=[kpoint adjoint,doubled]
\tikzstyle{dkpointadj}=[kpoint adjoint,doubled]
\tikzstyle{dkpointconj}=[kpoint conjugate,doubled]
\tikzstyle{dkpointtrans}=[kpoint transpose,doubled]

\tikzstyle{kscalar}=[kpoint common, shape=EBox, inner xsep=-1pt, inner ysep=3pt,font=\small]
\tikzstyle{kscalarconj}=[kpoint common, shape=WBox, inner xsep=-1pt, inner ysep=3pt,font=\small]

\tikzstyle{spekpoint}=[kpoint sc,minimum height=5mm,inner sep=3pt]
\tikzstyle{spekcopoint}=[kpoint adjoint sc,minimum height=5mm,inner sep=3pt]

\tikzstyle{dspekpoint}=[spekpoint,doubled]
\tikzstyle{dspekcopoint}=[spekcopoint,doubled]

 \tikzstyle{upground}=[circuit ee IEC,thick,ground,rotate=90,scale=2.5]
 \tikzstyle{downground}=[circuit ee IEC,thick,ground,rotate=-90,scale=2.5]
 \tikzstyle{bigground}=[regular polygon,regular polygon sides=3,draw=gray,scale=0.50,inner sep=-0.5pt,minimum width=10mm,fill=gray]

\tikzstyle{arrs}=[-latex,font=\small,auto]
\tikzstyle{arrow plain}=[arrs]
\tikzstyle{arrow dashed}=[dashed,arrs]
\tikzstyle{arrow bold}=[very thick,arrs]
\tikzstyle{arrow hide}=[draw=white!0,-]
\tikzstyle{arrow reverse}=[latex-]
\tikzstyle{cdnode}=[]

%% file: dynamics_channels_arxiv_v4.bbl
\begin{thebibliography}{999}

\bibitem[Hardy(2005)]{hardy2005probability}
Hardy, L.
 Probability theories with dynamic causal structure: A new framework
  for quantum gravity.
 {\em arXiv} {\bf 2005}, arXiv:gr-qc/0509120.

\bibitem[Hardy(2007)]{hardy2007towards}
Hardy, L.
 Towards quantum gravity: A framework for probabilistic theories with
  non-fixed causal structure.
 {\em J. Phys. Math. Theor.} {\bf 2007},
  {\em 40},~3081.

\bibitem[Markes and Hardy(2011)]{markes2011entropy}
Markes, S.; Hardy, L.
 Entropy for theories with indefinite causal {structure.} 
 \emph{J. Phys. Conf. Ser.} \textbf{2011}, \emph{306}, 012043.

\bibitem[Hardy(2009)]{hardy2009quantum}
Hardy, L.
 Quantum gravity computers: On the theory of computation with
  indefinite causal structure. In {\em Quantum Reality, Relativistic Causality,
  and Closing the Epistemic Circle}; Springer: {Dordrecht, The Netherlands,}  2009; pp. 379--401.

\bibitem[Chiribella \em{et~al.}(2009)Chiribella, D'Ariano, Perinotti, and
  Valiron]{chiribella2009beyond}
Chiribella, G.; D'Ariano, G.; Perinotti, P.; Valiron, B.
 Beyond quantum computers.
 {\em arXiv} {\bf 2009}, arXiv:0912.0195.

\bibitem[Chiribella \em{et~al.}(2013)Chiribella, D'Ariano, Perinotti, and
  Valiron]{qswitch}
Chiribella, G.; D'Ariano, G.M.; Perinotti, P.; Valiron, B.
 Quantum computations without definite causal structure.
 {\em Phys. Rev. A} {\bf 2013}, {\em 88},~022318.

\bibitem[Branciard(2016)]{branciard2016witnesses}
Branciard, C.
 Witnesses of causal nonseparability: An introduction and a few case
  studies.
 {\em Sci. Rep.} {\bf 2016}, {\em 6},~26018.

\bibitem[Ara{\'u}jo \em{et~al.}(2014)Ara{\'u}jo, Costa, and
  Brukner]{araujo2014computational}
Ara{\'u}jo, M.; Costa, F.; Brukner, {\v{C}}.
 Computational advantage from quantum-controlled ordering of gates.
 {\em Phys. Rev. Lett.} {\bf 2014}, {\em 113},~250402.

\bibitem[Feix \em{et~al.}(2015)Feix, Ara{\'u}jo, and Brukner]{feix2015quantum}
Feix, A.; Ara{\'u}jo, M.; Brukner, {\v{C}}.
 Quantum superposition of the order of parties as a communication
  resource.
 {\em Phys. Rev. A} {\bf 2015}, {\em 92},~052326.

\bibitem[Gu{\'e}rin \em{et~al.}(2016)Gu{\'e}rin, Feix, Ara{\'u}jo, and
  Brukner]{guerin2016exponential}
Gu{\'e}rin, P.A.; Feix, A.; Ara{\'u}jo, M.; Brukner, {\v{C}}.
 Exponential communication complexity advantage from quantum
  superposition of the direction of communication.
 {\em Phys. Rev. Lett.} {\bf 2016}, {\em 117},~100502.

\bibitem[Kristj{\'a}nsson \em{et~al.}(2020)Kristj{\'a}nsson, Chiribella, Salek,
  Ebler, and Wilson]{kristjansson2020resource}
Kristj{\'a}nsson, H.; Chiribella, G.; Salek, S.; Ebler, D.; Wilson, M.
 Resource theories of communication.
 {\em New J. Phys.} {\bf 2020}, {\emph{22}, 073014}.

\bibitem[Zhao \em{et~al.}(2020)Zhao, Yang, and Chiribella]{zhao2020quantum}
Zhao, X.; Yang, Y.; Chiribella, G.
 Quantum metrology with indefinite causal order.
 {\em Phys. Rev. Lett.} {\bf 2020}, {\em 124},~190503.

\bibitem[Oreshkov \em{et~al.}(2012)Oreshkov, Costa, and Brukner]{PM}
Oreshkov, O.; Costa, F.; Brukner, {\v{C}}.
 Quantum correlations with no causal order.
 {\em Nat. Commun.} {\bf 2012}, {\em 3},~1092.

\bibitem[Aharonov \em{et~al.}(1964)Aharonov, Bergmann, and
  Lebowitz]{aharonov1964time}
Aharonov, Y.; Bergmann, P.G.; Lebowitz, J.L.
 Time symmetry in the quantum process of measurement.
 {\em Phys. Rev.} {\bf 1964}, {\em 134},~B1410.

\bibitem[Silva \em{et~al.}(2017)Silva, Guryanova, Short, Skrzypczyk, Brunner,
  and Popescu]{silva2017connecting}
Silva, R.; Guryanova, Y.; Short, A.J.; Skrzypczyk, P.; Brunner, N.; Popescu, S.
 Connecting processes with indefinite causal order and multi-time
  quantum states.
 {\em New J. Phys.} {\bf 2017}, {\em 19},~103022.

\bibitem[Oreshkov and Cerf(2016)]{oreshkov2016operational}
Oreshkov, O.; Cerf, N.J.
 Operational quantum theory without predefined time.
 {\em New J. Phys.} {\bf 2016}, {\em 18},~073037.

\bibitem[Perinotti(2017)]{perinotti2017causal}
Perinotti, P.
 Causal structures and the classification of higher order quantum
  computations. In {\em Time in Physics}; Springer: {Cham, Switzerland}, 2017; pp. 103--127.

\bibitem[Kissinger and Uijlen(2017)]{kissinger2017categorical}
Kissinger, A.; Uijlen, S.
 A categorical semantics for causal structure.
 In Proceedings of the 2017 32nd Annual ACM/IEEE Symposium on Logic in
  Computer Science (LICS), {Reykjavik, Iceland, 20--23 June} {2017;} 
 pp. 1--12.

\bibitem[Bisio and Perinotti(2019)]{bisio2019theoretical}
Bisio, A.; Perinotti, P.
 Theoretical framework for higher-order quantum theory.
 {\em Proc. R. Soc. A} {\bf 2019}, {\em
  475},~20180706.

\bibitem[Chiribella \em{et~al.}(2008{\natexlab{a}})Chiribella, D'Ariano, and
  Perinotti]{chiribella2008quantum}
Chiribella, G.; D'Ariano, G.M.; Perinotti, P.
 Quantum circuit architecture.
 {\em Phys. Rev. Lett.} {\bf 2008}, {\em 101},~060401.

\bibitem[Chiribella \em{et~al.}(2008{\natexlab{b}})Chiribella, D'Ariano, and
  Perinotti]{chiribella2008transforming}
Chiribella, G.; D'Ariano, G.M.; Perinotti, P.
 Transforming quantum operations: Quantum supermaps.
 {\em EPL (Europhys. Lett.)} {\bf 2008}, {\em 83},~30004.

\bibitem[Chiribella \em{et~al.}(2010)Chiribella, D’Ariano, and Perinotti]{chiribella2010probabilistic}
Chiribella, G.; D’Ariano, G.M.; Perinotti, P.
 Probabilistic theories with purification.
 {\em Phys. Rev. A---At. Mol. Opt. Phys.} {\bf 2010}, {\em 81},~062348.

\bibitem[Castro-Ruiz \em{et~al.}(2018)Castro-Ruiz, Giacomini, and
  Brukner]{castro2018dynamics}
Castro-Ruiz, E.; Giacomini, F.; Brukner, {\v{C}}.
 Dynamics of quantum causal structures.
 {\em Phys. Rev. X} {\bf 2018}, {\em 8},~011047.

\bibitem[Jia \em{et~al.}(2018)Jia, Sakharwade, et~al.]{jia2018tensor}
Jia, D.; Sakharwade, N.
 Tensor products of process matrices with indefinite causal structure.
 {\em Phys. Rev. A} {\bf 2018}, {\em 97},~032110.

\bibitem[Gu{\'e}rin \em{et~al.}(2019)Gu{\'e}rin, Krumm, Budroni, and
  Brukner]{guerin2019composition}
Gu{\'e}rin, P.A.; Krumm, M.; Budroni, C.; Brukner, C.
 Composition rules for quantum processes: A no-go theorem.
 {\em New J. Phys.} {\bf 2019}, {\emph{21}, 012001}.

\bibitem[Horodecki \em{et~al.}(2009)Horodecki, Horodecki, Horodecki, and
  Horodecki]{horodecki2009quantum}
Horodecki, R.; Horodecki, P.; Horodecki, M.; Horodecki, K.
 Quantum entanglement.
 {\em Rev. Mod. Phys.} {\bf 2009}, {\em 81},~865.

\bibitem[Castro-Ruiz \em{et~al.}(2020)Castro-Ruiz, Giacomini, and
  Brukner]{castro2020comment}
Castro-Ruiz, E.; Giacomini, F.; Brukner, {\v{C}}.
 Comment on" Revisiting dynamics of quantum causal structures--when
  can causal order evolve?''.
 {\em arXiv} {\bf 2020}, arXiv:2009.10027.

\bibitem[Banaszek \em{et~al.}(2004)Banaszek, Dragan, Wasilewski, and
  Radzewicz]{Banaszek_et_al_2004}
Banaszek, K.; Dragan, A.; Wasilewski, W.; Radzewicz, C.
 Dynamics of quantum causal structures.
 {\em Phys. Rev. Lett.} {\bf 2004}, {\em 92},~257901.

\bibitem[Procopio \em{et~al.}(2015)Procopio, Moqanaki, Ara{\'u}jo, Costa,
  Calafell, Dowd, Hamel, Rozema, Brukner, and
  Walther]{procopio2015experimental}
Procopio, L.M.; Moqanaki, A.; Ara{\'u}jo, M.; Costa, F.; Calafell, I.A.; Dowd,
  E.G.; Hamel, D.R.; Rozema, L.A.; Brukner, {\v{C}}.; Walther, P.
 Experimental superposition of orders of quantum gates.
 {\em Nat. Commun.} {\bf 2015}, {\em 6},~7913.

\bibitem[Rubino \em{et~al.}(2017)Rubino, Rozema, Feix, Ara{\'u}jo, Zeuner,
  Procopio, Brukner, and Walther]{rubino2017experimental}
Rubino, G.; Rozema, L.A.; Feix, A.; Ara{\'u}jo, M.; Zeuner, J.M.; Procopio,
  L.M.; Brukner, {\v{C}}.; Walther, P.
 Experimental verification of an indefinite causal order.
 {\em Sci. Adv.} {\bf 2017}, {\em 3},~e1602589.

\bibitem[Goswami \em{et~al.}(2018)Goswami, Giarmatzi, Kewming, Costa,
  Branciard, Romero, and White]{goswami2018indefinite}
Goswami, K.; Giarmatzi, C.; Kewming, M.; Costa, F.; Branciard, C.; Romero, J.;
  White, A.
 Indefinite causal order in a quantum switch.
 {\em Phys. Rev. Lett.} {\bf 2018}, {\em 121},~090503.

\bibitem[Stinespring(1955)]{dilation}
Stinespring, W.F.
 Positive functions on {$C^*$}-algebras.
 {\em Proc. Am. Math. Soc.} {\bf 1955}, {\em 6},~211--216.

\bibitem[Banaszek \em{et~al.}(2019)Banaszek, Dragan, Wasilewski, and
  Radzewicz]{Cywinski_et_al_2019}
Banaszek, K.; Dragan, A.; Wasilewski, W.; Radzewicz, C.
 How to detect qubit-environment entanglement generated during qubit
  dephasing.
 {\em Phys. Rev. A} {\bf 2019}, {\em 100},~022318.

\bibitem[Hardy(2001)]{hardy2001quantum}
Hardy, L.
 Quantum theory from five reasonable axioms.
 {\em arXiv} {\bf 2001}, arXiv:quant-ph/0101012.

\bibitem[Dakic and Brukner(2011)]{dakic2011quantum}
Dakic, B.; Brukner, C.
 Quantum Theory and Beyond: Is Entanglement Special? In \emph{Deep Beauty}; Halvorson, H., Ed.; {Cambridge University Press: Cambridge, UK}, 2011.

\bibitem[Masanes and M{\"u}ller(2011)]{masanes2011derivation}
Masanes, L.; M{\"u}ller, M.P.
 A derivation of quantum theory from physical requirements.
 {\em New J. Phys.} {\bf 2011}, {\em 13},~063001.

\bibitem[Ara{\'u}jo \em{et~al.}(2017)Ara{\'u}jo, Feix, Navascu{\'e}s, and
  Brukner]{araujo2017purification}
Ara{\'u}jo, M.; Feix, A.; Navascu{\'e}s, M.; Brukner, {\v{C}}.
 A purification postulate for quantum mechanics with indefinite causal
  order.
 {\em Quantum} {\bf 2017}, {\em 1},~10.


\end{thebibliography}
